\documentclass[a4paper,english,nofootinbib]{revtex4}
\usepackage[T1]{fontenc}
\usepackage[latin9]{inputenc}
\usepackage{textcomp}
\usepackage{amstext}
\usepackage{amssymb}
\usepackage{graphicx}

\makeatletter

\pdfpageheight\paperheight
\pdfpagewidth\paperwidth

\DeclareRobustCommand{\greektext}{%
  \fontencoding{LGR}\selectfont\def\encodingdefault{LGR}}
\DeclareRobustCommand{\textgreek}[1]{\leavevmode{\greektext #1}}
\DeclareFontEncoding{LGR}{}{}
\DeclareTextSymbol{\~}{LGR}{126}
\newcommand{\lyxmathsym}[1]{\ifmmode\begingroup\def\b@ld{bold}
  \text{\ifx\math@version\b@ld\bfseries\fi#1}\endgroup\else#1\fi}

\providecommand{\tabularnewline}{\\}

\@ifundefined{textcolor}{}
{%
 \definecolor{BLACK}{gray}{0}
 \definecolor{WHITE}{gray}{1}
 \definecolor{RED}{rgb}{1,0,0}
 \definecolor{GREEN}{rgb}{0,1,0}
 \definecolor{BLUE}{rgb}{0,0,1}
 \definecolor{CYAN}{cmyk}{1,0,0,0}
 \definecolor{MAGENTA}{cmyk}{0,1,0,0}
 \definecolor{YELLOW}{cmyk}{0,0,1,0}
 }

\makeatother

\usepackage{babel}
\begin{document}

\title{Oblique Parameters and Extra Generations via OPUCEM}

\author{Ece A\c{s}\i{}lar}

\email{ece.asilar@cern.ch}

\affiliation{Middle East Technical University, Physics Department, Ankara, Turkey}

\author{Esin Çavlan}

\email{echavlan@gmail.com}

\affiliation{Afyon Kocatepe University, Physics Department, Afyon, Turkey}

\author{Oktay Do\u{g}angün}

\email{oktay.dogangun@cern.ch}

\affiliation{University of Naples \& INFN, Department of Physical Sciences, Naples,
Italy}

\author{Sinan Kefeli }

\email{sinan.kefeli@boun.edu.tr}

\author{V. Erkcan Özcan}

\email{Erkcan.Ozcan@cern.ch}

\affiliation{Bo\u{g}aziçi University, Physics Department, Bebek, Istanbul, Turkey}

\author{Mehmet \c{S}ahin}

\email{mehmet.sahin@usak.edu.tr}

\affiliation{Usak University, Department of Physics, Usak, Turkey}

\author{Gökhan Ünel}

\email{Gokhan.Unel@cern.ch}

\affiliation{University of California at Irvine, Department of Physics and Astronomy,
Irvine, USA }
\begin{abstract}
Recent improvements to OPUCEM, the tool for calculation of the contributions
of various models to oblique parameters, are presented. OPUCEM is
used to calculate the available parameter space for the four family
Standard Model given the current electroweak precision data. It is
shown that even with the restrictions on Higgs boson and new quark
masses presented in the 2011 Autumn conferences, there is still enough
space to allow a fourth generation with Dirac type neutrinos. For
Majorana type neutrinos, the allowed parameter space is even larger.
The electroweak precision data also appear to favor non-zero mixing
between light and fourth generations, thus effectively reducing the
current experimental limits on the masses of the new quarks, which
assume that the mixing with the third generation is dominant. Additionally,
 disregarding the lack of a clear Higgs signal from the LHC and focusing
only an electroweak precision data comptability, calculations with
OPUCEM show that, the existing electroweak data are compatible with
the presence of a 5th and also a 6th generation in certain regions
of the parameter space.
\end{abstract}
\maketitle

\section{introduction }

The categorization of the electroweak (EW) corrections based on their
contribution types dates back to a study of photon propagated four-fermion
processes \cite{stu-orig}. In the original nomenclature, the corrections
to vertices, box diagrams and bremsstrahlung diagrams were all considered
as {}``\emph{direct}'' whereas the propagator corrections due to
vacuum polarization effects were all named as {}``\emph{oblique}''
since these participate to the computations in an indirect manner.
An extended review of this approach and its application of the methodology
to Beyond the Standard Model (BSM) theories helped coining the term
oblique parameters \cite{stu-def} usually represented by the letters
\emph{$S$, $T$} and $U$. The Standard Model is defined by the values
$S=T=U=0$ for a given top quark and Higgs boson mass. In a BSM theory
with new fermion doublets, the $S$ parameter estimates the size of
the additional fermion sector and the $T$ parameter measures the
isospin symmetry violation in that sector. In such a model, the \emph{$U$}
parameter is dependent on W boson width, thus insensitive to new physics. 

As the one-loop exact calculations for a model are tedious and error
prone, an open-source C/C$++$ library called OPUCEM (Oblique Parameters
Using C with Error-checking Machinery) was introduced to improve the
reliability and reproducibility of computations in scientific publications
\cite{opucemWeb}. This library provides functions to calculate the
contributions to the oblique parameters from a number of BSM models.
A command line program constitutes an example on how to use the library
functions and a graphical user interface facilitates the library's
use by non-programmers. Additionally, tools for plotting error ellipses
in the $S-T$ plane and extensive documentation are provided. Among
the physics models currently implemented, one can cite Standard Model
with four fermion families (SM4)\cite{6li} and 2 Higgs Doublet Models
(2HDM). Formulas for computing the effect of adding new fermion doublets
with Majorana-type neutrinos and of the mixing between the quark generations
are also available. The oblique parameters $S$, $T$ and $U$ are
computed by using both exact one-loop calculations and with some well-defined
approximations for various models, providing an additional machinery
for error checking, beyond the various internal self-crosscheck mechanisms.

Various studies of the SM4 mass and mixing parameters performed using
OPUCEM were previously discussed in$~$\cite{opucem}. This note describes
the recent additions to the library, and to the command line and graphical
tools, as present in the OPUCEM version 00-00-07. The overall available
parameter space in the SM4 model and the compatibilities of BSM theories
with further additional generations are discussed. Despite the recent
negative results from the Large Hadron Collider (LHC) in the search
for heavy quarks using about 1$\,$fb$^{-1}$ of data$~$\cite{LHCquarkSearches},
it is shown that there is still available room in the SM4 parameter
space. The improvements to the OPUCEM package since the last publication
\cite{opucem}, can be summarized as follows: (i) the addition of
the mixing in the quark sector between the new heavy quarks and the
second generation quarks, namely the function calls to calculate $T$
value in the presence of the parameter in addition to the previously
available $\theta{}_{34}$; (ii) machinery to perform parameter scans
through the command line interface, in order to facilitate the exploration
of the parameter space defined by the fermion and Higgs masses and
the quark mixings;  (iii) extension of the graphical user interface
to include the fifth and sixth generation fermion masses for both
Dirac and Majorana type neutrinos. In the current implementation,
if the masses of the \emph{$N{}^{th}$} generation fermions are set
to zero, the calculations are automatically reduced to those for a
model of $N-1$ generations.

\section{OPUCEM on the SM4}

\begin{figure}[h]
\begin{centering}
\includegraphics[scale=0.5]{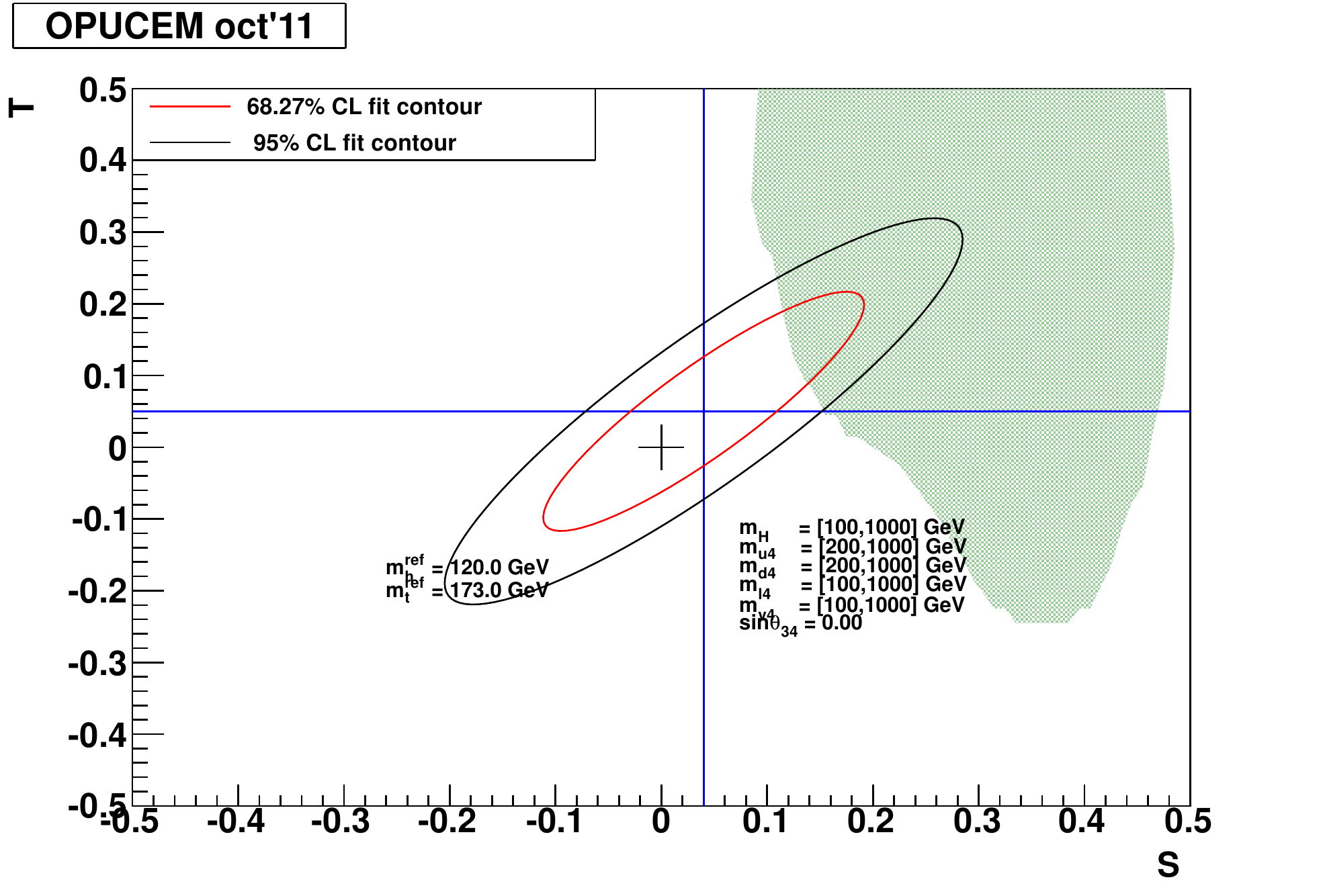}
\par\end{centering}

\caption{A parameter scan of the SM4 model with OPUCEM using Gfitter default
values and parameters. The plot has been produced in a format that
allows direct comparison with \cite{Gfitter}.\label{fig:Opucem_Gfitter}}
\end{figure}

OPUCEM and Gfitter are the two open-source tools available for the
calculation of the oblique parameters for a number of models%
\footnote{After the first version of this paper appeared on the arXiv, the authors
were informed of the G4LHC package$~$\cite{g4lhc} that includes
the software for the SM4 $S$, $T$ calculations that were performed
in$~$\cite{kribs}.%
} \cite{Gfitter}. In addition to SM4, Gfitter's theory repository
includes other models such as Extra Dimensions and Little Higgs Models.
However for the SM4 model, Gfitter does not incorporate the effect
of the mixing between the quarks of the fourth and other generations,
and it is limited only to Dirac type neutrinos. While these two tools
have been created for different purposes, the fact that two recent
implementations of the $S$, $T$ formulas exist provides an opportunity
to check their reliability. As the original goal of the OPUCEM library
has been to provide well-tested and error-free results (as tested
multiple times before), a comparison between the results of Gfitter
and OPUCEM has been performed. For such a comparison, the input reference
values and electroweak fit results should be the same in both packages.
Since the Gfitter results are presented with two different sets of
input values (with $U=0$ forced and $U=free$), the reference values
and the definitions of the error ellipses for the two programs are
given in Table$~$\ref{tab:Reference-values} for the readers convenience. The
default values in OPUCEM have been obtained from the LEP Electroweak
Working Group through private communication (summer 2009 results)
and from the Tevatron Electroweak Working Group \cite{topmass}. Figure$~$\ref{fig:Opucem_Gfitter}
shows a scan of the fourth generation parameter space performed with
OPUCEM using the Gfitter reference values for the $U=free$ case.
This figure can directly be compared to the Gfitter fourth generation
parameter scan results as presented in$~$\cite{Gfitter}, as it has
been prepared with the same scan and axis ranges for the new fermion
and Higgs boson masses (as indicated on the figure itself), and with
only Dirac-type neutrinos and no mixings between generations. The
green area is the region of the $S$, $T$ plane that is accessible
by SM4, as the input parameters are scanned through. Red and black
ellipses are the 68\% and 95\% CL contours of the experimentally allowed
$S$, $T$ values. The plus sign shows the SM reference point where
all the oblique parameters vanish, whereas the blue horizontal and
vertical lines define the center of the error ellipses given in the
first row of Table \ref{tab:Reference-values}. Comparison of this
plot with the Figure$~$12 of \cite{Gfitter} shows perfect agreement
between the results of the two software packages.

\begin{table}[h]
\caption{Reference values to use for comparison between Gfitter and OPUCEM
results. $\rho$ indicates the correlation coefficient between the
$S$ and $T$ values.\label{tab:Reference-values}}

\begin{tabular}{|c|c|c|c|c|c|c|}
\hline 
 & $m_{H}$ & $m_{top}$ & $S$ & $T$ & $U$ & $\rho$\tabularnewline
\hline 
\hline 
Gfitter 1 & 120 & 173 & 0.04\textpm{}0.10 & 0.05\textpm{}0.11 & 0.08\textpm{}0.11 & 0.89\tabularnewline
\hline 
Gfitter 2 & 120 & 173 & 0.07\textpm{}0.09 & 0.10\textpm{}0.08 & 0 & 0.88\tabularnewline
\hline 
OPUCEM & 115 & 173.1 & 0.07\textpm{}0.10  & 0.1067\textpm{}0.09 & 0 & 0.85 \tabularnewline
\hline 
\end{tabular}
\end{table}

For the rest of this section, the Standard Model studies with four
generations of fermions (SM4) is discussed so as to determine the
effect of the EW precision data on $S-T$ parameter space. Cases with
Dirac and Majorana type neutrinos, both stable and unstable, are considered.
The aim is to perform a parameter space scan to investigate the viability
of the SM4 model, given the current status of the searches and the
electroweak precision data results. The initial values of mass parameters
originate from the direct searches, usually summarized by the Particle
Data Group (PDG)\cite{PDG} and updated with publications from current
direct search experiments. For the final values a common mass limit
of 1 TeV, not too far from the partial wave unitarity bound, seems
to be preferred choice in the literature. Unless stated otherwise,
the step size in the mass scans of the $S-T$ plane is taken to be
10 GeV which is a good compromise between the execution speed and
the accuracy of the results. For the remainder of this work, OPUCEM
defaults shown in the last row of Table \ref{tab:Reference-values}
are used in $S-T$ parameter space plots.

\subsection{With Dirac type neutrinos}

\begin{figure}[h]
\noindent \includegraphics[width=0.48\textwidth]{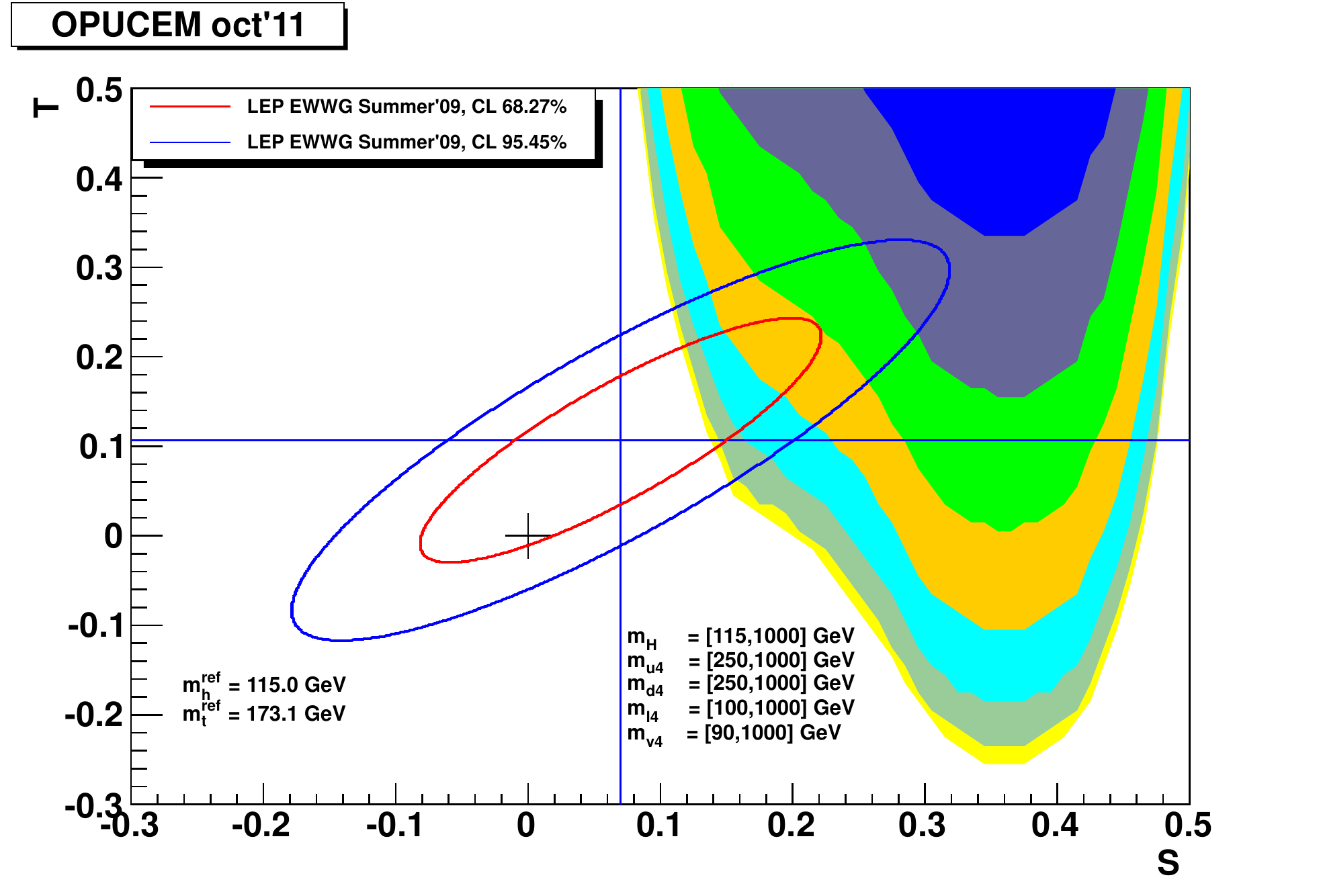}\includegraphics[width=0.48\textwidth]{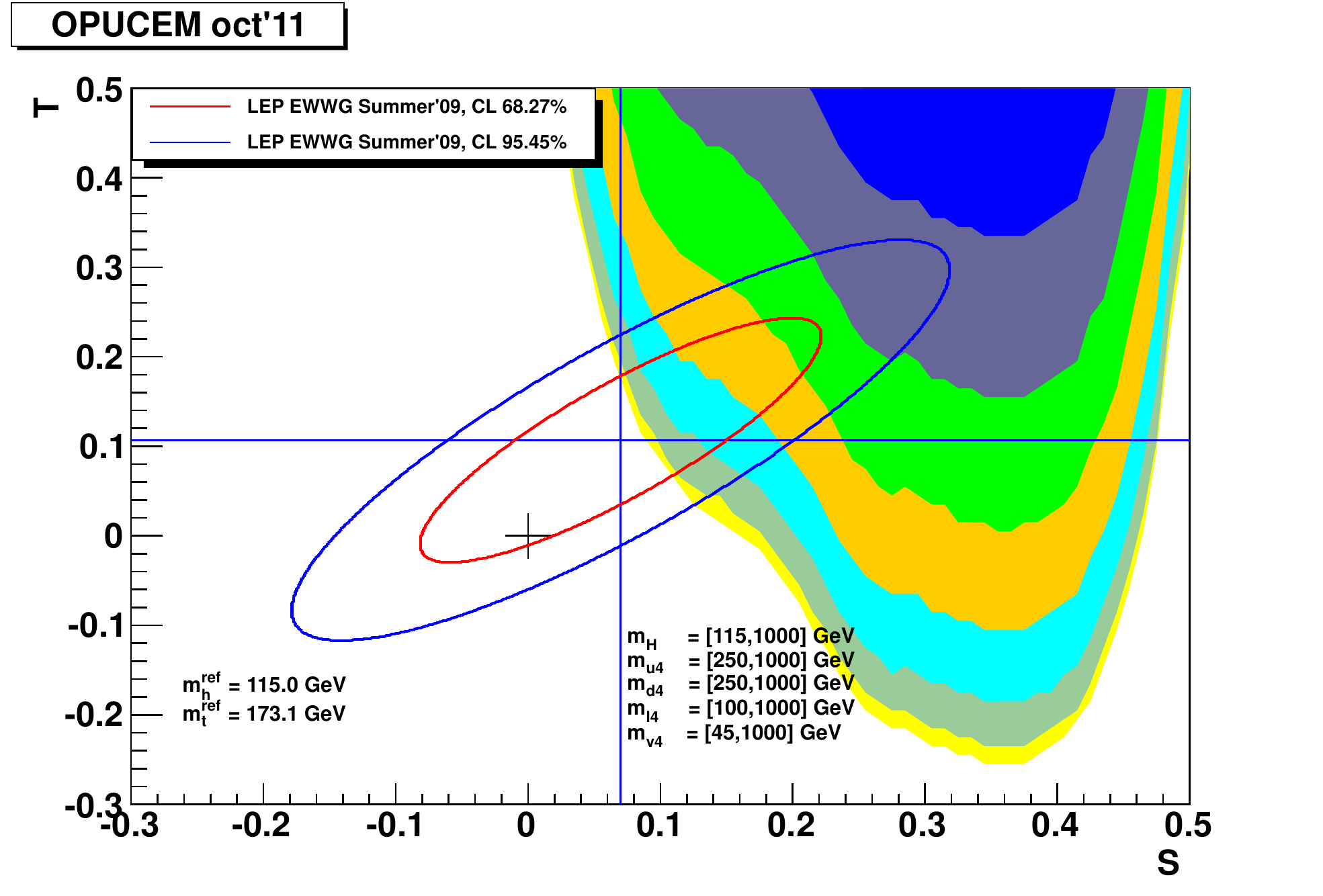}\caption{Impact of the mixing between third and fourth family quarks on the
fourth generation $S$, $T$ space for \emph{unstable} (left) and
\emph{stable} (right) Dirac-type neutrinos obtained using the PDG
mass limit values. The color-filled regions show the set of points
in the plane that is reachable by varying all the fourth generation
parameters except the mixing angles. Colors correspond to $|\sin$$\theta{}_{34}|$
varying between $0.0$ and $0.6$.\label{fig:Mixing-angle-dependence_pdg}}
\end{figure}

For a fourth generation with a Dirac type neutrino, 6 parameters are
commonly considered: masses of the new fermions, the Higgs boson mass
and the sine of the mixing angle between the third and fourth generation
quarks. The mixing between third and fourth generation quarks, $\theta{}_{34}$
, plays an important role in the determination of the best fit to
the EW precision data. Left-hand side of Figure$~$\ref{fig:Mixing-angle-dependence_pdg}
shows the impact of this mixing on the allowed parameter space when
the PDG mass limits on the new fermions mass values are considered
for an unstable Dirac neutrino. In this scenario, the initial default
mass pattern is set as: $m$$_{u4}=m_{d4}=250$ GeV, $m$$_{H}=115$
GeV, $m$$_{\ell4}=100$ GeV and $m$$_{v4}=90$ GeV. Seven different
mixing angle values are considered to investigate the effect on the
$S-T$ plane: $|\sin\theta{}_{34}|=0.0$, $0.1$, $0.2$, $0.3$,
$0.4$, $0.5$, $0.6$. In Figure$~$\ref{fig:Mixing-angle-dependence_pdg},
each scan with a different mixing angle is shown as a contour with
a different color, the largest contour (hence allowed parameter space)
in the $S-T$ plane corresponds to no-mixing and the smallest contour,
shown with dark blue, corresponds to $|\sin\theta{}_{34}|=0.6$. As seen
from the intersection of these colored contours with the $S$, $T$
ellipses extracted from the EW precision data, the size of the allowed
region gets smaller with increasing mixing, such that in order to
be compatible with the EW precision data at the $2\sigma$ level,
the mixing between the third and fourth generation quarks has to have
an upper limit, $|\sin\theta{}_{34}|\lesssim0.6$, independent of
the other parameters. In general, small mixing angles are preferred
and so as to keep the fourth generation in the $1\sigma$ error ellipse,
the mixing angle should satisfy $|\sin\theta{}_{34}|<0.4$.

Since the minimum experimentally allowed mass for a stable neutrino differs
from the unstable case, the parameter space is re-scanned for the
stable Dirac case. As seen on the right-hand side of Figure$~$\ref{fig:Mixing-angle-dependence_pdg},
the decrease in the neutrino mass limit downs to $45$$\,$GeV reflects
itself as an expansion of the allowed parameter space with respect
to the previous case. However, the constraints for the mixing angle
remain as before.

\begin{figure}[h]
\includegraphics[width=0.48\textwidth]{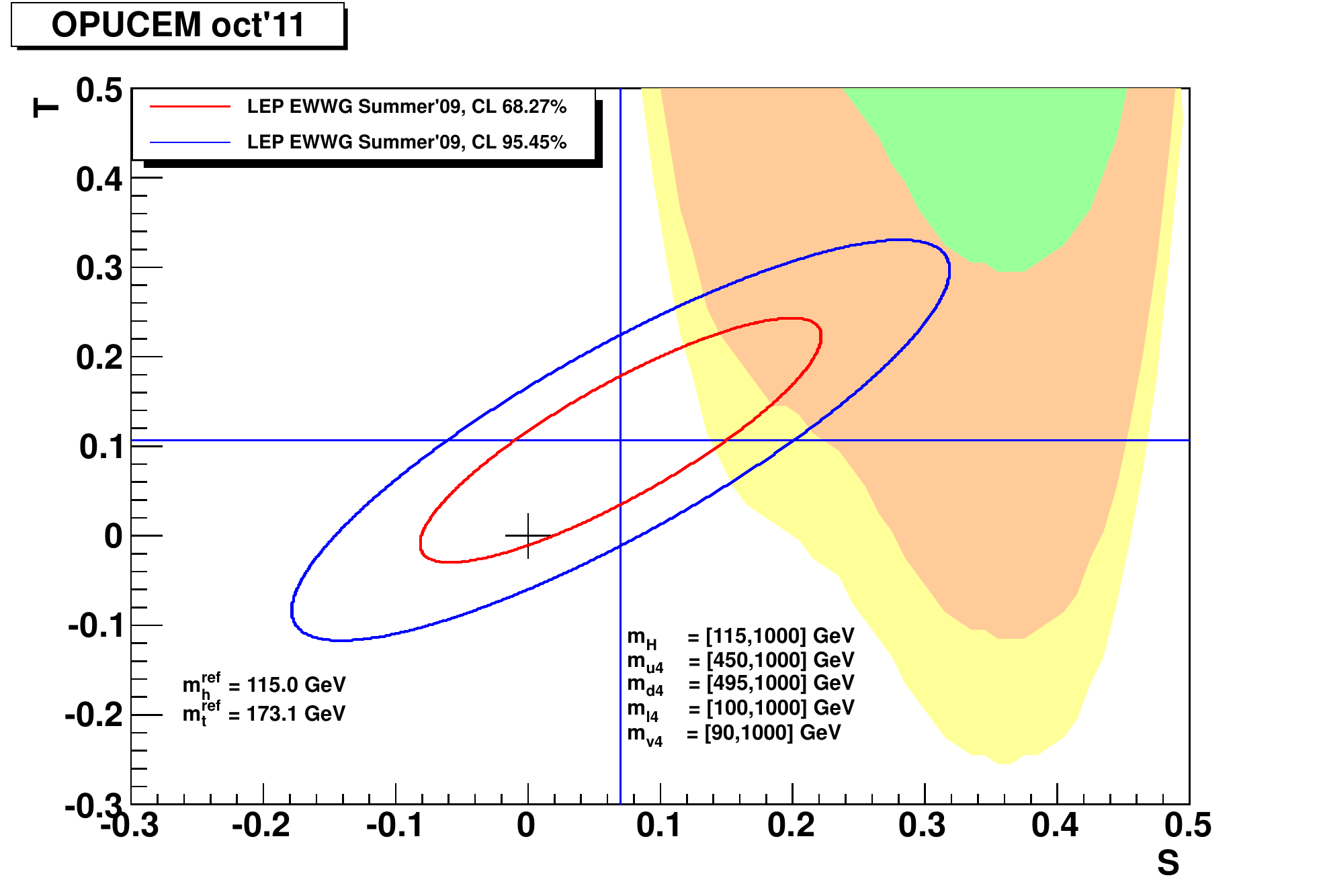}\includegraphics[width=0.48\textwidth]{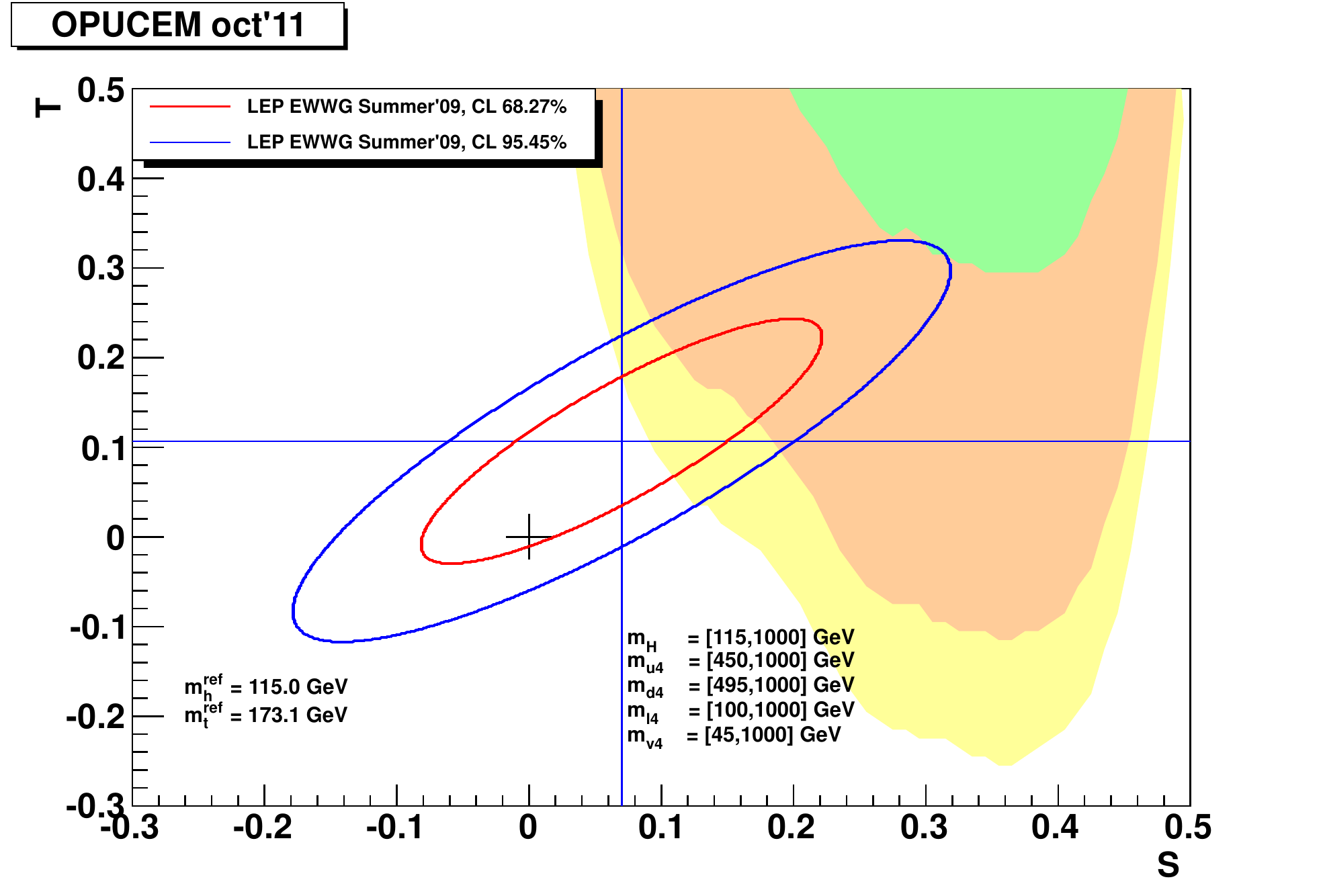}\caption{Mixing angle dependence of the fourth generation quarks. $S$, $T$
parameter space for \emph{unstable} (left) and \emph{stable} (right)
Dirac-type neutrinos using the updated mass limit values from the
LHC: $m_{u4}>450$ GeV and $m_{d4}>495$ GeV. The 3 colored regions
correspond to $|\sin$$\theta{}_{34}|$$=0.0$, $0.1$ and $0.2$
(yellow, orange and green, respectively). \label{fig:Mixing-angle-dependence_LHC}}
\end{figure}

In Autumn 2011, the CMS Collaboration reported new limits on the minimum
allowed heavy quark mass values corresponding to $m_{u4}>450$ GeV
and $m_{d4}>495$ GeV at 95\% CL\cite{KaiFC,CMS2,CMS3}. The parameter
space after this update is shown in Figure$~$\ref{fig:Mixing-angle-dependence_LHC}.
The colored contours correspond to three different mixing angle values
$|\sin\theta{}_{34}|=0.0$, $0.1$ and $0.2$. In this scenario, due
to the reduction of the allowed parameter space, the 2$\sigma$ error
ellipse corresponds to a maximum value of the mixing angle, $|\sin\theta{}_{34}|<0.2$,
both for stable and unstable neutrinos.

\begin{figure}[h]
\includegraphics[scale=0.5]{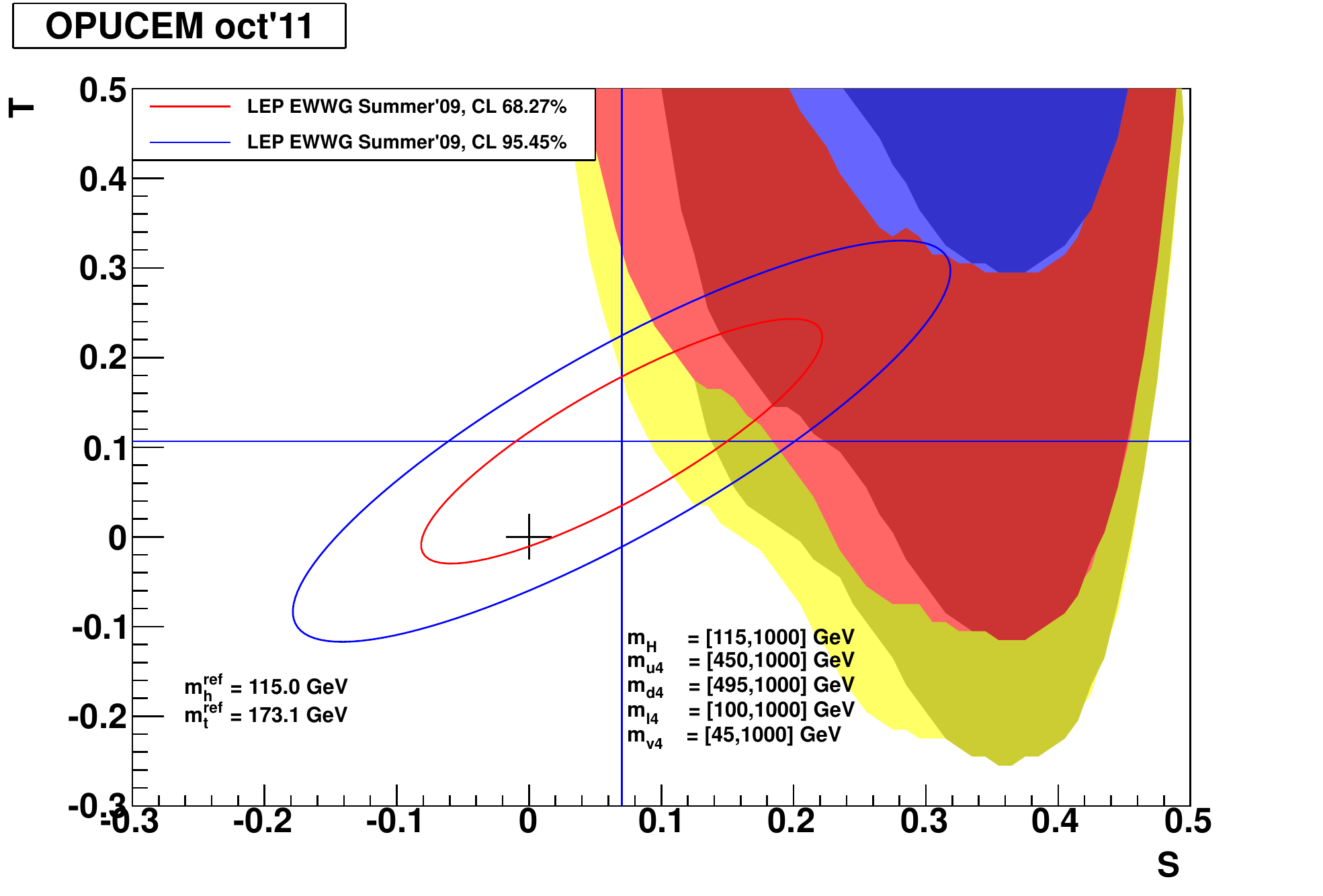}\caption{Mixing angle dependence of stable vs unstable neutrinos with updated
mass limit values: $m_{u4}>450$ GeV and $m_{d4}>495$ GeV. \label{fig:Mixing-angle-dependence_quasi.stable_vs_unstable}}
\end{figure}

The impact of assuming a stable neutrino is better illustrated in
Figure \ref{fig:Mixing-angle-dependence_quasi.stable_vs_unstable}
where three different mixing angle values for stable and unstable
neutrinos are investigated. The considered mixing angle values are
$|\sin$$\theta{}_{34}|$$=0.0$, $0.1$ and $0.2$ where the allowed
regions size decreases as the mixing angle increases. In the plot,
darker colors represent the unstable Dirac neutrino case and light
colors represent the stable case. One can see from the figure that
the stable Dirac type neutrino slightly enlarges the available parameter
space in the negative $S$ direction, independently of the mixing
angle.

\begin{figure}[h]
\includegraphics[scale=0.5]{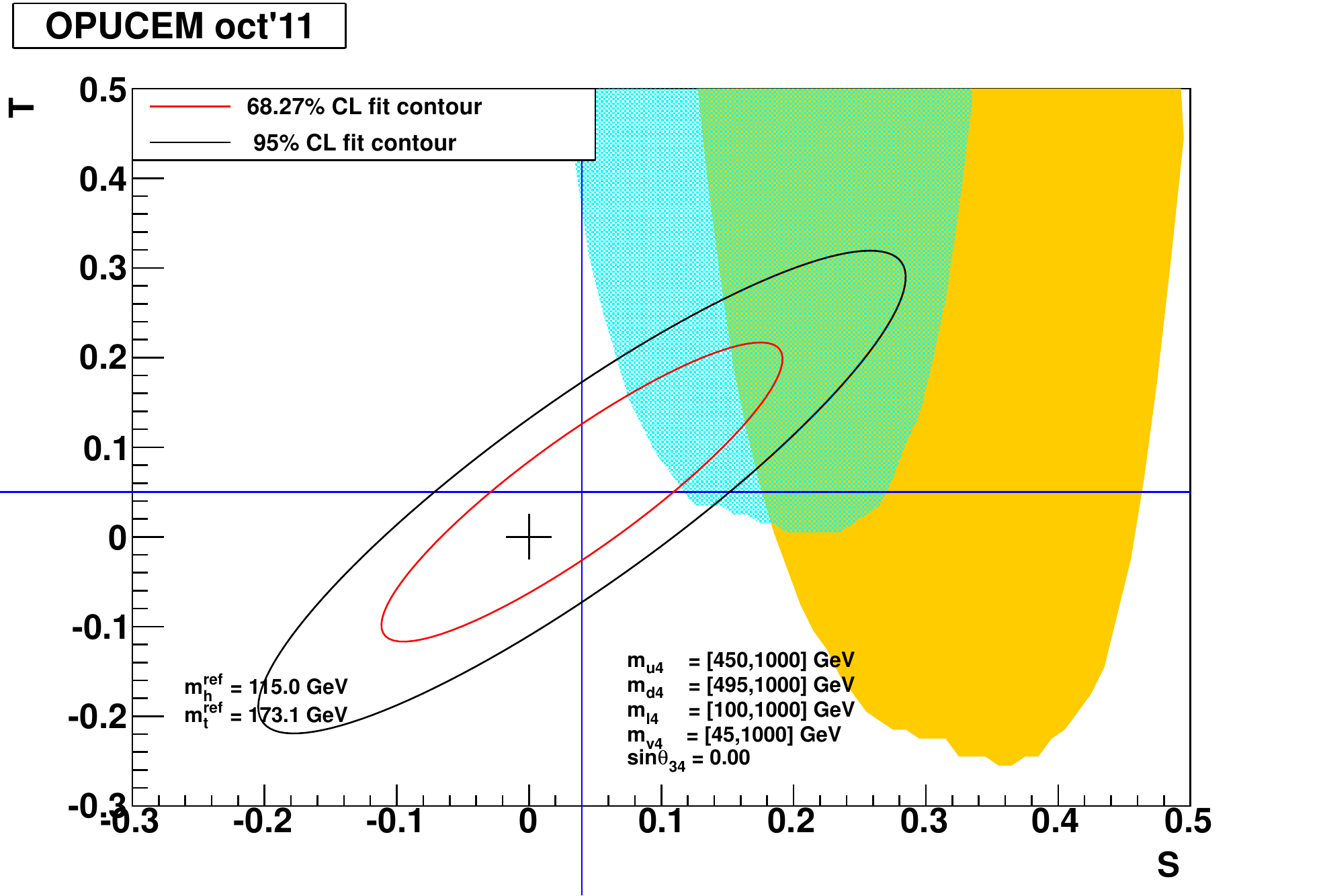}\caption{Light Higgs mass (light blue) and heavy Higgs mass (orange) dependence
of the fourth family fermions in \emph{stable} Dirac neutrino case
with updated mass limit values: $m_{u4}>450$$\,$GeV and $m_{d4}>495$$\,$GeV.\label{fig:Light-Higgs_Heavy-Higgs_dependence_Summer2011_stable}}
\end{figure}

Results from Summer 2011 which references the CMS outcomes have also
excluded a SM Higgs boson in the mass range $120<m_{H}<600$ GeV with
95\% CL \cite{CMS4,ATLAS}. The remaining mass windows for the Higgs
boson are for a heavy one of mass larger than 600$\,$GeV and a light
one between 115 and 120$\,$GeV. Figure$~$\ref{fig:Light-Higgs_Heavy-Higgs_dependence_Summer2011_stable}
indicates these states for the Dirac stable neutrino case with $|\sin\theta{}_{34}|=0.0$.
The light (heavy) Higgs boson scenario is shown with light blue (orange).
One can see that light Higgs boson is favored over a heavy one, but
even a heavy Higgs boson is still compatible with a fourth generation
according to the EW precision data.

\subsection{With Majorana type neutrinos}

\begin{figure}[h]
\includegraphics[width=0.48\textwidth]{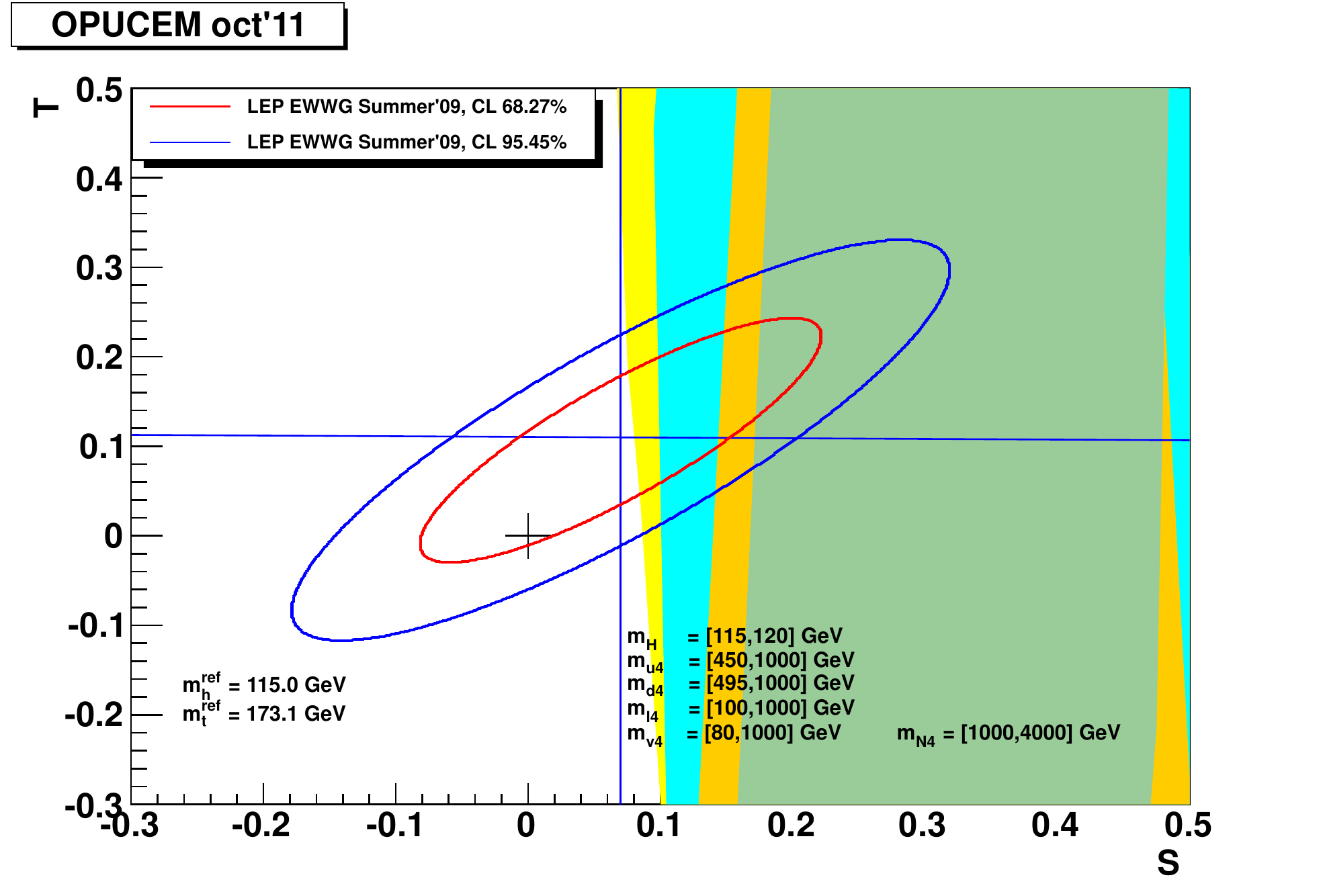}
\includegraphics[width=0.48\textwidth]{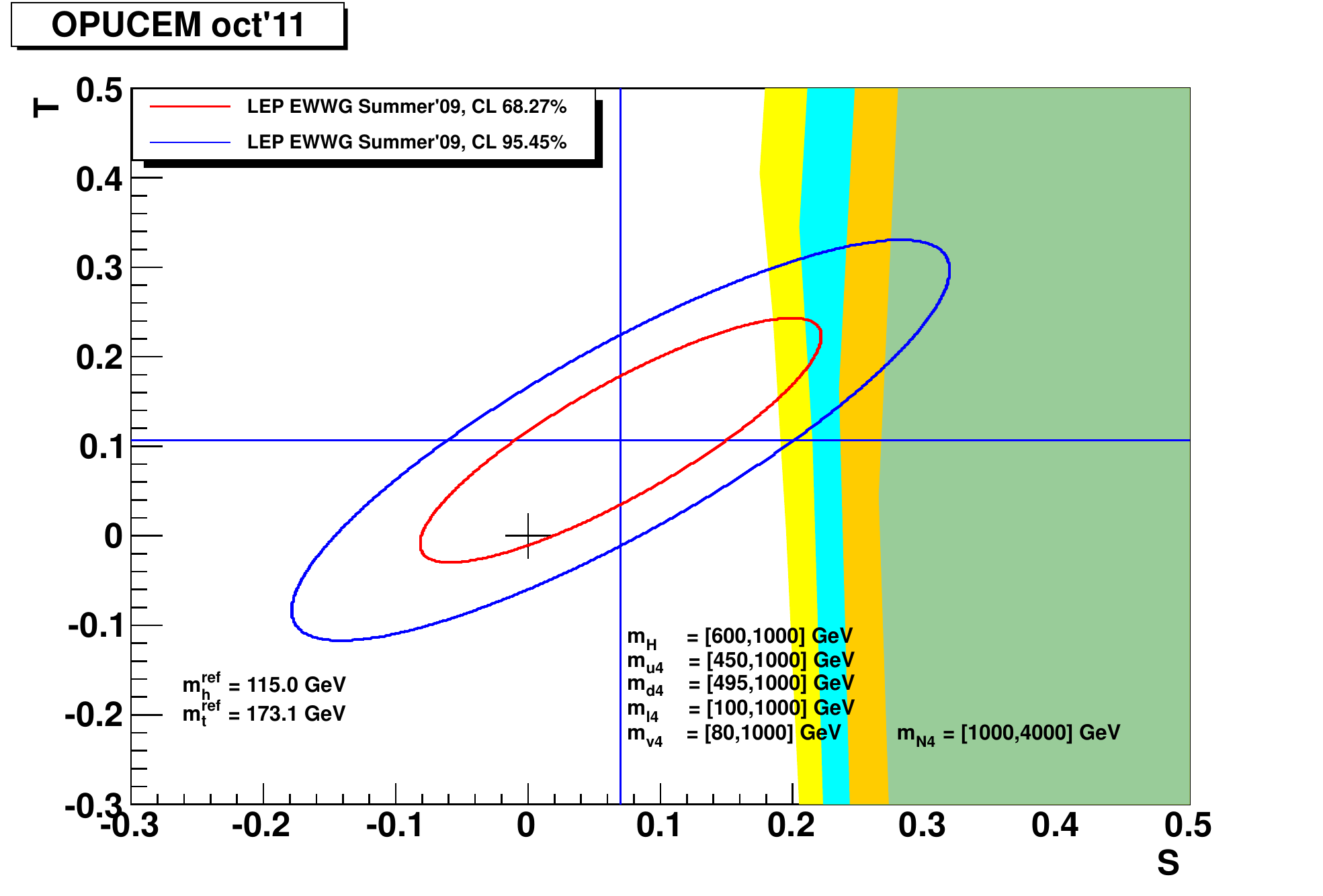}\caption{Light (left) and heavy (right) Higgs mass dependence for the third
and fourth family fermion in \emph{unstable} Majorana case in terms
of different mixing angle values. \label{fig:Heavy_Light_Higgs_mass-mixing_angle_dependence_Summer2011_majorana}The
colored regions correspond to |$\sin$$\theta$| = 0.0, 0.2, 0.4 and
0.6 (Different colored contours represent the allowed parameter space
are yellow, cyan, orange and green respectively).}
\end{figure}

A fourth generation with a Majorana type neutrino requires an extra
mass parameter, namely the mass of the heavier neutrino denoted by
$m{}_{N4}$. Therefore, seven parameters have been considered for
the oblique parameter calculations: masses of the new fermions including
the mass of the heavier neutrino, the Higgs boson mass and the sine
of the mixing angle between third and fourth generation quarks. Figure$~$\ref{fig:Heavy_Light_Higgs_mass-mixing_angle_dependence_Summer2011_majorana}
contains a parameter scan for the unstable neutrino case, considering
both light (left) and heavy (right) Higgs boson cases. The initial
fermion mass pattern is set as: $m$$_{u4}=450$ GeV, $m$$_{d4}=495$
GeV, $m$$_{H}=115$ GeV, $m$$_{\ell4}=100$ GeV, $m$$_{N4}=1000$
GeV and $m$$_{\nu4}=80$ GeV, where the quark mass lower limits are
mostly from the CMS results \cite{CMS1,CMS2,CMS3}. The maximum values
of the masses have been taken as 1 TeV for all fermions except for
the heavier neutrino where a maximum of 4 TeV has been considered.
The light Higgs mass is tested for two values, 115$\,$GeV and 120$\,$GeV,
and the heavy Higgs mass is scanned from 600$\,$GeV to 990$\,$GeV
in steps of 30$\,$GeV. The mixing angle values under consideration
are $|\sin\theta{}_{34}|=0.0$, $0.2$, $0.4$ and $0.6$ represented
by contours of different colors: yellow, cyan, orange and green respectively.
A simple comparison of the two plots reveals that the light Higgs
boson is still the preferred scenario, just like it was the case for
the Dirac type neutrinos. In this scenario, although smaller mixing
angles are preferred, the highest considered value of $|\sin\theta{}_{34}|=0.6$
can still be tuned to appear in the 1$\sigma$ error ellipse. On the
other hand, for the heavy Higgs bosons case, which is less favored,
the highest value of the mixing angle allowed by the 1$\sigma$ error
ellipse is $|\sin$$\theta{}_{34}|$$=$ $0.2$. It can also be concluded
that for the Majorana type neutrinos, the additional parameter ($m{}_{N4}$)
gives more flexibility to the choice of parameters. Therefore Majorana
neutrino case provides a wider allowed parameter region than Dirac
neutrino case. Especially by tuning the ratio of the masses of the
new neutrinos, it is possible to make some previously unaccessible
portions of the parameter space, compatible with the EW precision
data. As a final remark, stable Majorana case was also separately
considered but not included in this section. As the only difference
between the two cases is the decrease of the neutrino mass lower limit
down to 40 GeV, all conclusions for the unstable neutrino scenario
remain also valid for the stable neutrino case, with an increase of
the allowed parameter space size.

\subsection{Fourth Generation mixing with light generations}

In the literature most experimental results are reported assuming
that the fourth generation quarks mix dominantly with the third generation,
\emph{i.e.} $BF(u_{4}\rightarrow Wb)=100\%$ and likewise for the
$d_{4}$. This choice is partially motivated by the gains of using
$b$-tagging in the data analysis. It is clear that a proper interpretation
of the direct quark search limits requires consideration of the mixings
with light generations. Furthermore, as has recently been shown in$~$\cite{lenzPaper},
taking quark mixings into account is extremely important in analyzing
the EW precision data as it can significantly change the favored mass
space. 

\begin{figure}[h]
\includegraphics[scale=0.4]{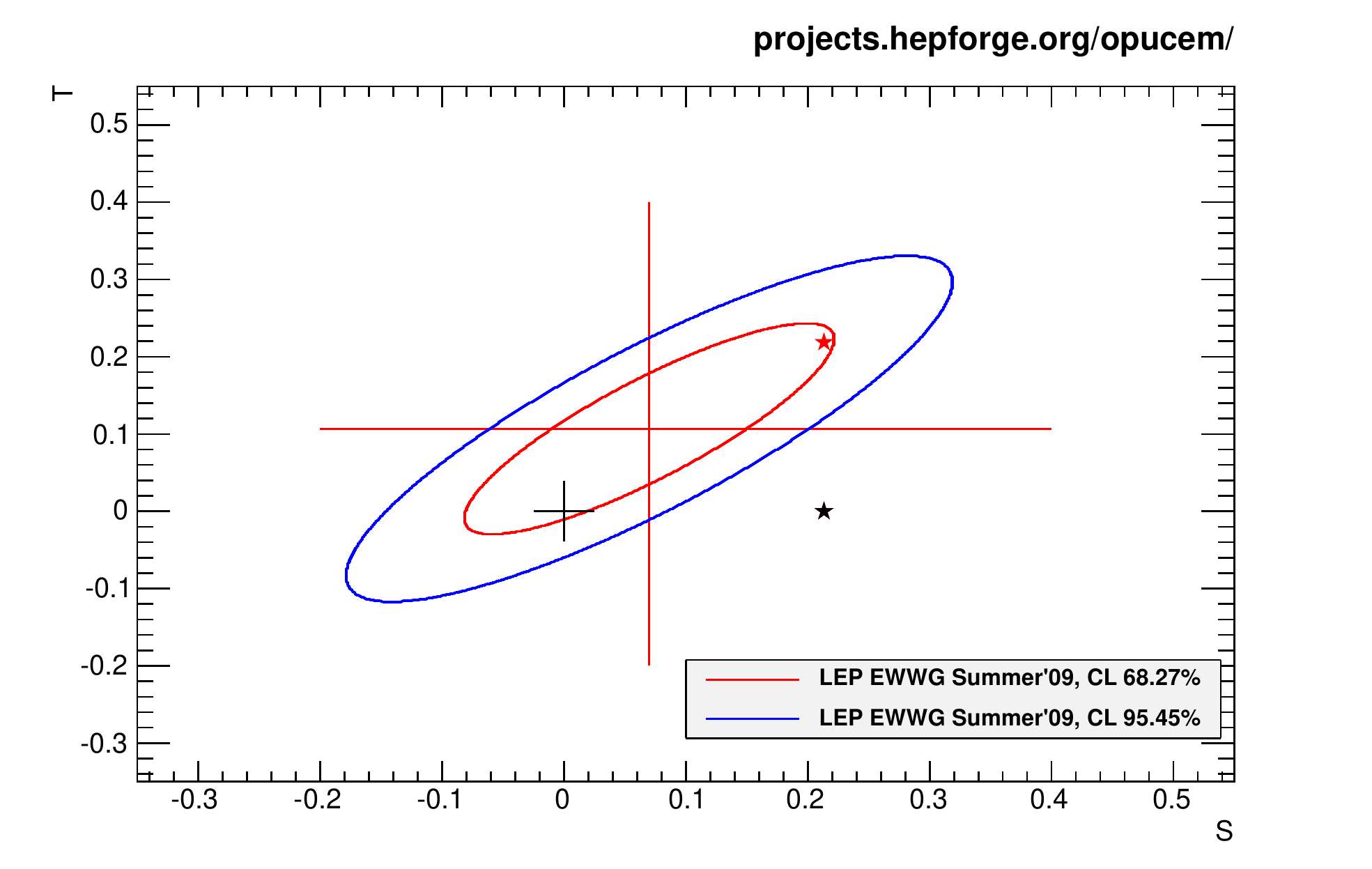} \caption{$S-T$ plane is drawn using OPUCEM for the fully degenerate case with
the masses of all the fourth generation fermions set to 400$\,$GeV.
While with no mixing such a scenario is essentially ruled out ($\lyxmathsym{\textgreek{D}}\lyxmathsym{\textgreek{q}}^{2}=22.1$,
black point), with mixings of $|\sin$$\theta_{24}|$ = 0.07, $|\sin$$\theta_{34}|$
= 0.14, the $\Delta\chi^{2}=2.0$ (red point), which is within 1$\lyxmathsym{\textgreek{sv}}$
error ellipse. \label{fig:Opucem_fourth_family_test}}
\end{figure}

OPUCEM has been used to test the implications of the quarks mixings
in the case of fully degenerate fourth generation fermions, which
is claimed to have been ruled out by the EW data according to the
PDG$~$\cite{PDG}. As seen in Figure$~$\ref{fig:Opucem_fourth_family_test},
while the fully degenerate case without mixings is indeed ruled out
with more than 99.99\% CL, it is possible to reach better than 1$\sigma$
agreement with EW precision data for certain choices of the mixing
parameters (in agreement with the findings of$~$\cite{lenzPaper})
One should note that since heavy Higgs boson masses tend to increase
the $S$ parameter while decreasing $T$, the introduction of the
mixings, which increase the $T$ parameter alone, can open up a significant
portion of the parameter space in general.

Therefore it is worth repeating some of the parameter scans detailed
in the earlier sections taking into account the mixings with the light
generations. The goal chosen for this section however, is to provide
some input to the interpretation of the LHC quark search limits, by
setting the mass parameters to certain example values and exploring
the favored values of the mixing parameters. To this end, the values
of the mixings allowed by the current measurements of the $3\times3$
CKM matrix elements have been determined. It is worth noting that
a full treatment of $4\times4$ CKM with EW constraints is important
for determining the overall allowed parameter space, as discussed
in$~$\cite{lenzPaper}. However, to perform first order estimations,
$3\times3$ CKM matrix measurements from the PDG$~$\cite{PDG}, and
imposing the unitarity condition of the $4\times4$ CKM matrix was
sufficient to determine the values of $K_{i4}\equiv|V_{i4}|^{2}$,
$i=u,c,t$. Uncertainties on $K_{i4}$, assumed to be Gaussian, can
be obtained by simple error propagation. This procedure yields: $K_{14}=0.0001\pm0.0006$,
$K_{24}=-0.10\pm0.07$, $K_{34}=0.22\pm0.12$.

\begin{figure}[h]
\centering{}\includegraphics[width=0.31\textwidth]{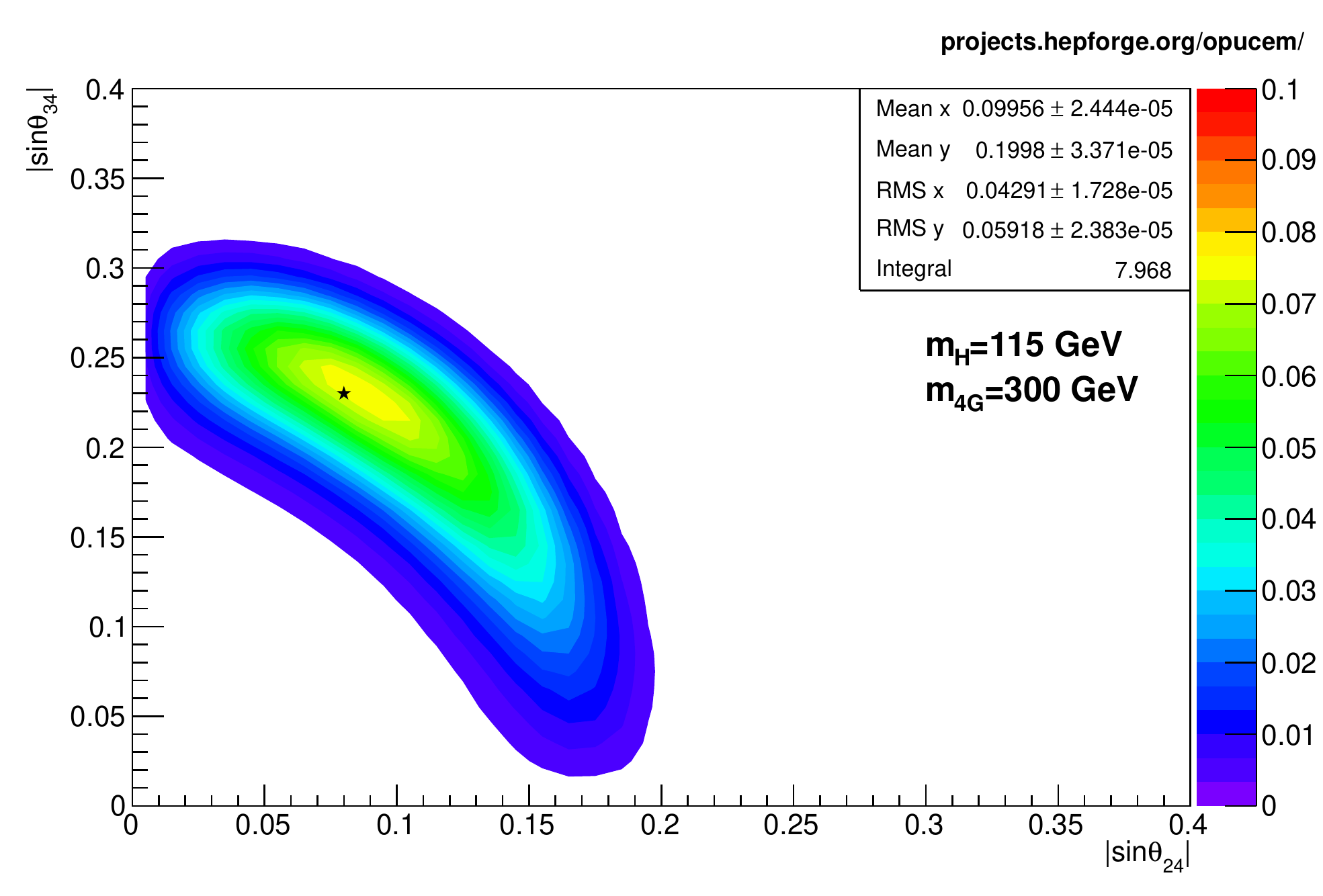}\includegraphics[width=0.31\textwidth]{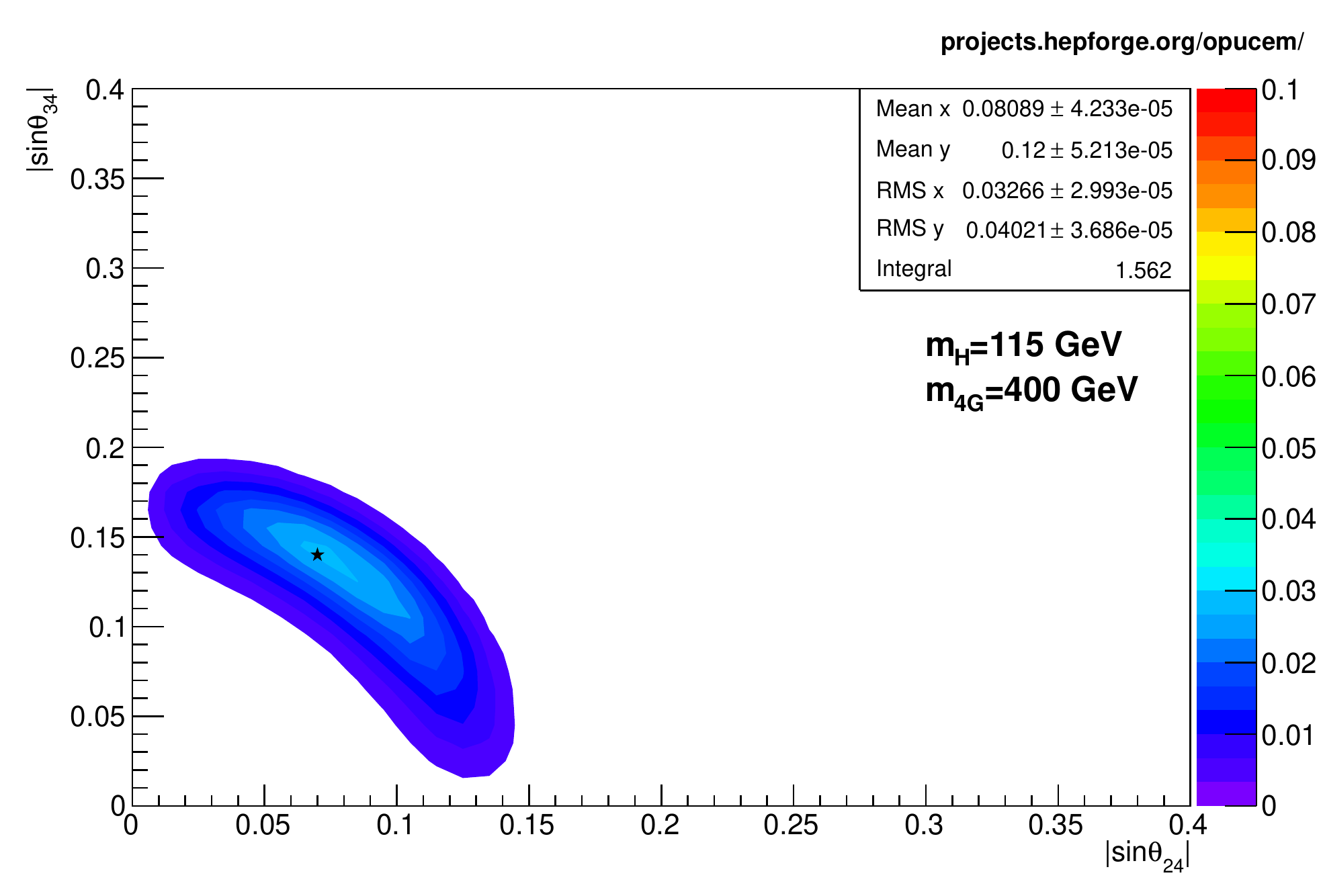}\includegraphics[width=0.31\textwidth]{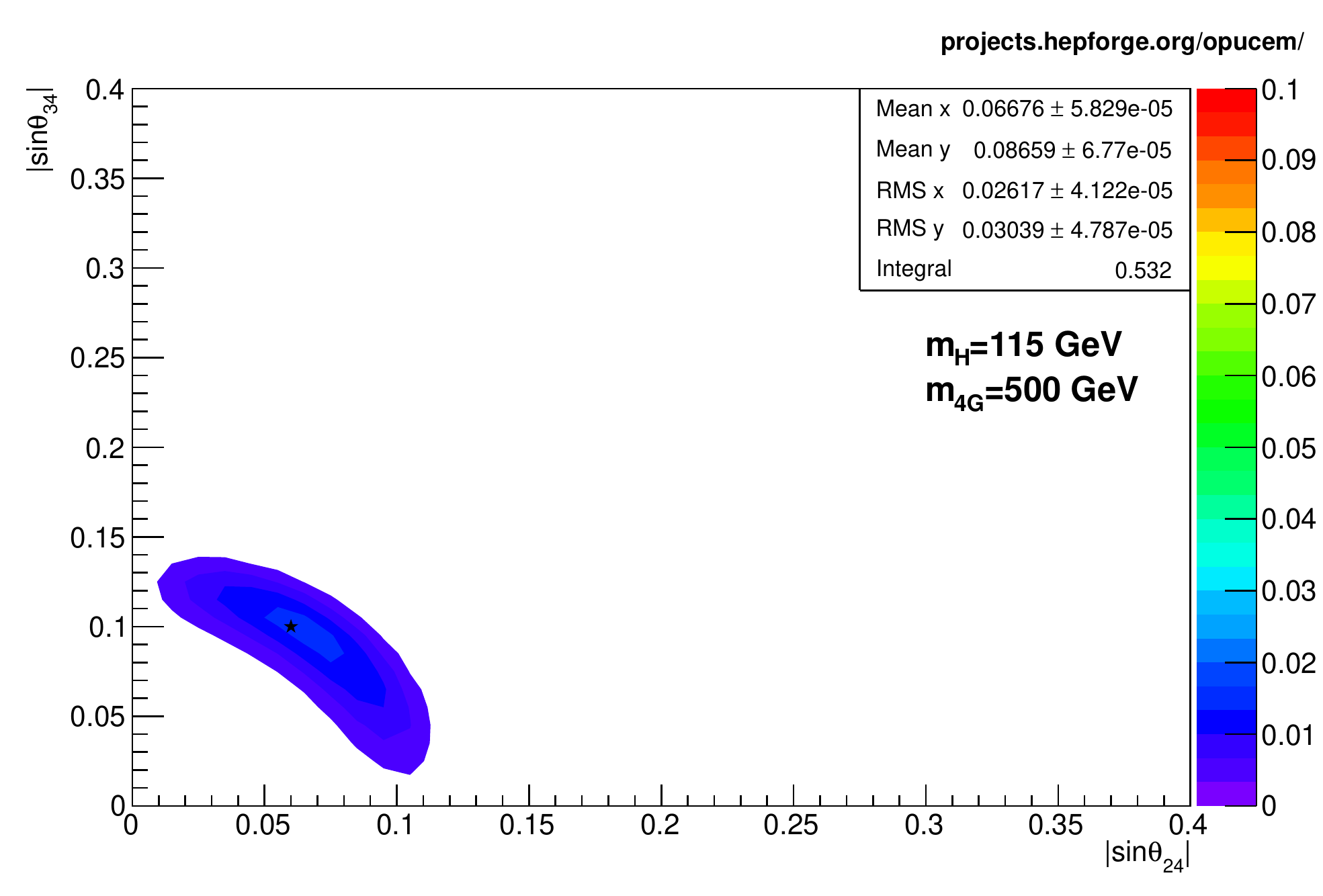}
\caption{2D histograms of $|\sin\theta_{24}|$ and $|\sin\theta_{34}|$, obtained
by using $\lyxmathsym{\textgreek{q}}^{2}$ probability in the $S$-$T$
space as weights, for three different values of the degenerate fourth-generation
fermion mass. (From left to right: $m_{\mathrm{4G}}=300$$\,$GeV,
$m_{\mathrm{4G}}=400$$\,$GeV, $m_{\mathrm{4G}}=500$$\,$GeV.) The
small black stars at the center of the colored contours indicate the
most favored $|\sin\theta_{24}|$ \textendash{} $|\sin\theta_{34}|$
pairs. The scale of the color-axis and the total integral of the weights
are normalized to $10^{3}$ non-negative $(K_{24}$, $K_{34})$ pairs.\label{fig:degenerate_fourth_hist2d_s34_s24}}
\end{figure}

Noting that the allowed space for the magnitude of $K_{14}$ is much
smaller than the others, the focus should be on the 2-4 and 3-4 mixings.
A large number (${\cal O}(10^{8})$) of Gaussian-distributed random
$(K_{24}$, $K_{34})$ pairs were generated using the center values
and uncertainties computed above. For each generated non-negative
$K_{24}$ and $K_{34}$, $S$, $T$ and $\lyxmathsym{\textgreek{D}}\lyxmathsym{\textgreek{q}}^{2}$
from the center of the $S$-$T$ ellipse were computed, for a Higgs
mass of 115$\,$GeV and assuming the neutrinos to be of Dirac type.
Then using $\lyxmathsym{\textgreek{q}}^{2}$ probability as weights,
a two dimensional histogram was filled with $|\sin\theta_{24}|$ and
$|\sin\theta_{34}|$ values. This procedure was repeated for a number
of different fourth-generation fermion masses ($m_{\mathrm{4G}}$)
to obtain plots of the mixing space as shown in Figure$~$\ref{fig:degenerate_fourth_hist2d_s34_s24}. 

\begin{figure}[h]
\includegraphics[width=0.48\textwidth]{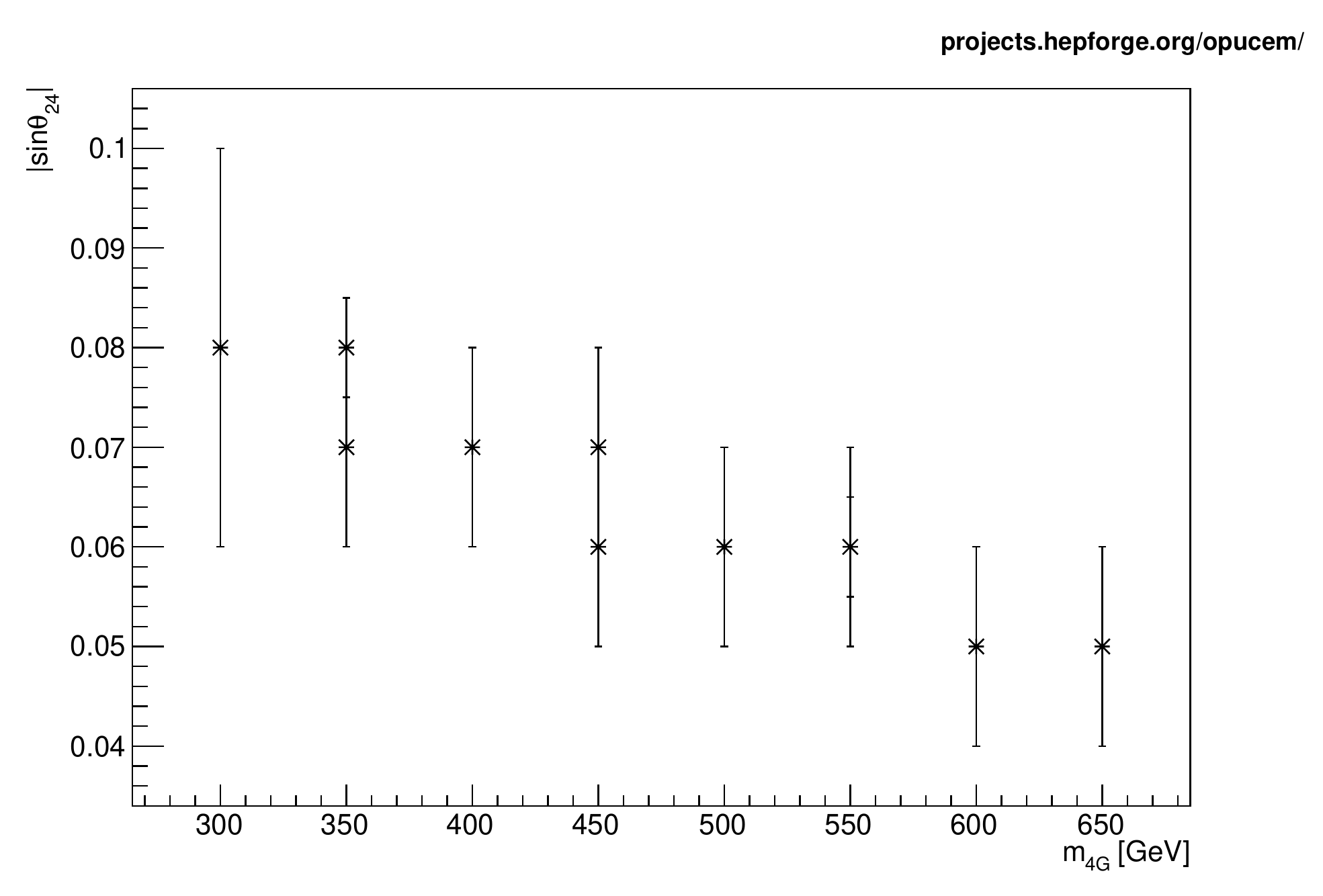}\includegraphics[width=0.48\textwidth]{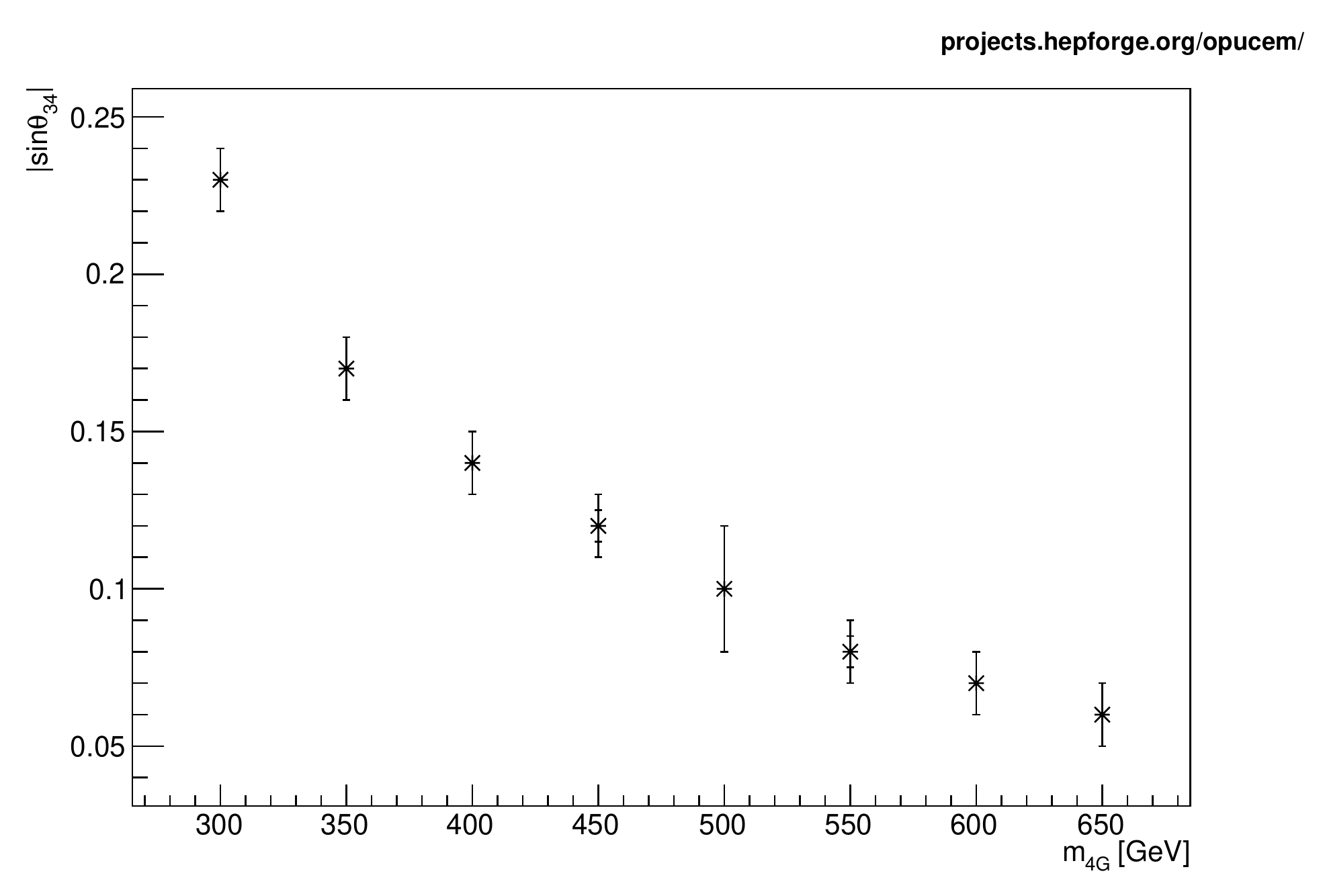}
\caption{Favored values of $|\sin\theta_{24}|$ (left) and $|\sin\theta_{34}|$
(right) as a function of fourth generation degenerate mass.\label{fig:deg4_s34_s24_f_m4g}}
\end{figure}

\newcommand{\slfrac}[2]{\left.#1\right/#2}By searching for the peak
in these 2D histograms, the most favored $(|\sin\theta_{24}|$, $|\sin\theta_{34}|)$
pair is determined as a function of $m_{\mathrm{4G}}$. One can observe
that the favored $\theta_{24}$ decreases slightly as $m_{\mathrm{4G}}$
increases, whereas $\theta_{34}$ shows a much more rapid decrease.
It is important to note that even at low masses, favored $\theta_{24}$
is not negligible compared to $\theta_{34}$ as shown in Figure$~$\ref{fig:deg4_s34_s24_f_m4g}.
It is also worth mentioning that the favored values are not far from
the results of the preliminary analysis presented in$~$\cite{lenzPresentation},
despite relying on much more limited experimental data. Since the
branching fraction for the decays of the fourth generation quarks
into final states involving third generation generation is approximately
given by $BF\sim\slfrac{|V_{tb'}|^{2}}(|V_{tb'}|^{2}+|V_{cb'}|^{2})$,
it is possible to extract the favored value of the $BF$ as well.
As seen in Figure$~$\ref{fig:Favored-BFs} left plot, this value
varies between 90\% and 60\% as a monotonically-decreasing function
of $m_{\mathrm{4G}}$.

\begin{figure}[h]
\centering{}\includegraphics[width=0.45\textwidth]{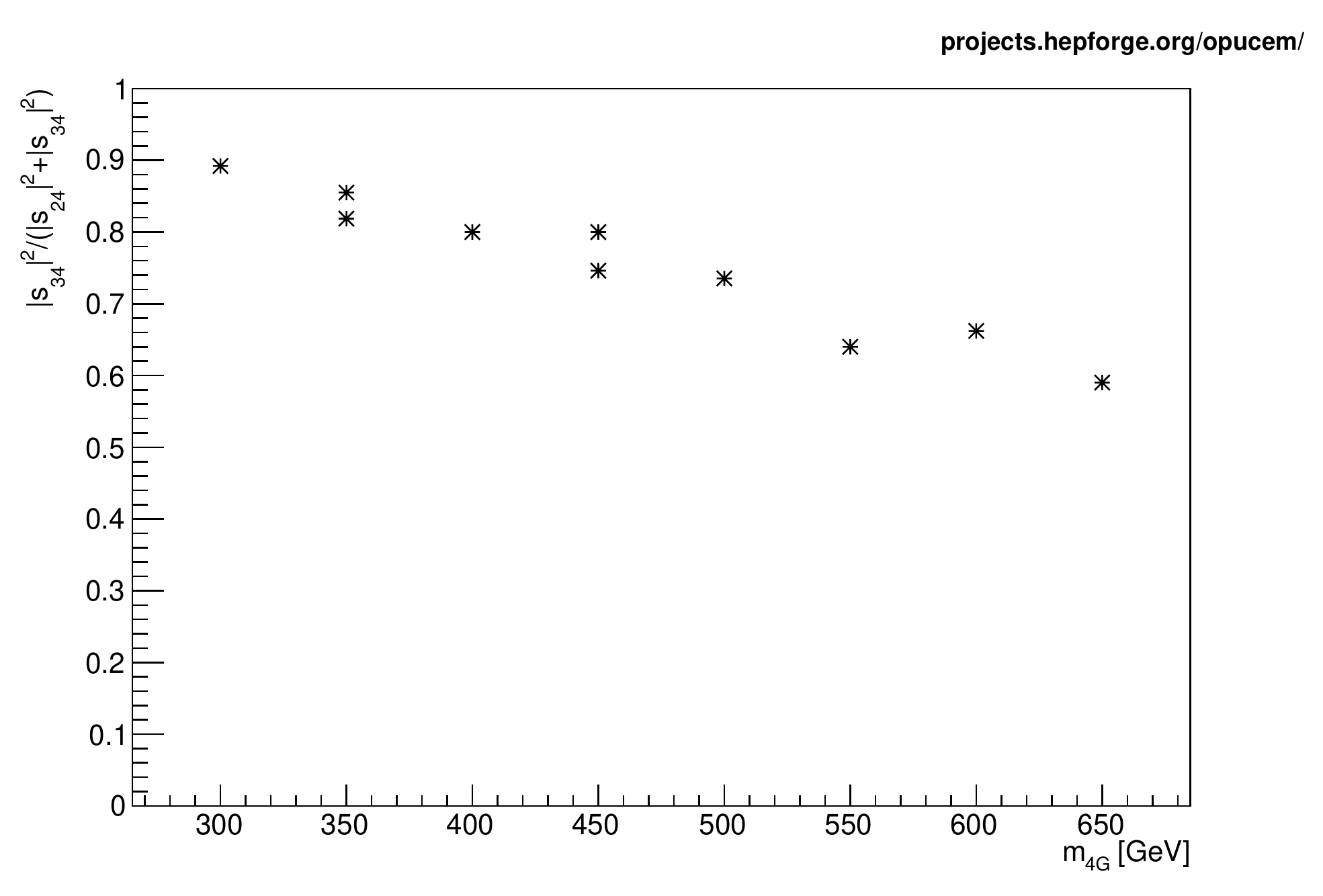}\includegraphics[width=0.45\textwidth]{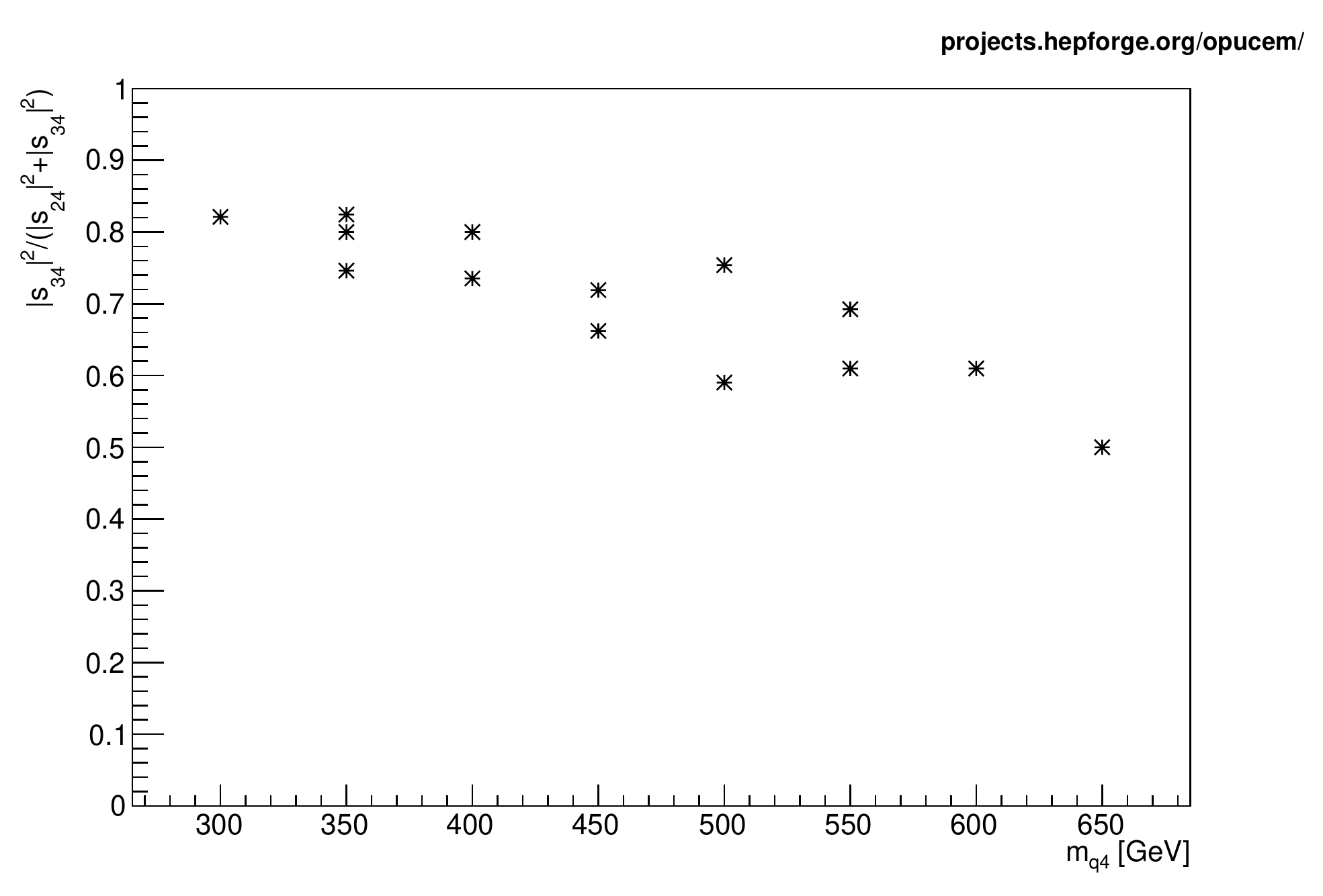}
\caption{Left: favored value of $BF\sim\slfrac{|V_{tb'}|^{2}}(|V_{tb'}|^{2}+|V_{cb'}|^{2})$
as a function of the fourth generation degenerate mass. This plot
has been prepared a number of times by iterating the full procedure
outlined in the text. For a couple of the $m_{\mathrm{4G}}$ values,
we extract slightly different $BF$ values from these iterations.
As a measure of the size of the systematic effects in our procedure,
we plot all such values. Right: same plot for the case of a non-degenerate
fourth generation scenario with only degenerate quark masses ($m_{q4}=m_{u4}=m_{d4}$).\label{fig:Favored-BFs}}
\end{figure}

An extensive analysis of the scenario with non-degenerate fourth generation
fermion masses requires the scan of a large parameter space. Therefore,
in order to gain some insight, a particular benchmark case is considered
in which a light Higgs boson decays into a pair of fourth generation
neutrinos, avoiding the current Higgs search limits from the LHC experiments$~$\cite{atlascmsHiggs,lightHnunu}.
The example benchmark point is taken to be: $m_{H}=130$$\,$GeV,
$m_{\lyxmathsym{\textgreek{n}}4}=60$$\,$GeV, $m_{\ell4}=120$$\,$GeV
and vary $m_{u4}$ = $m_{d4}$ between 300$\,$GeV and 650$\,$GeV.
Such a benchmark case is itself plausible without mixing, particularly
when $m_{\ell4}-m_{\nu4}$ mass difference is large. Therefore the
favored values of the mixings come out lower than the values in the
degenerate-generation case. However, the ratio of the squared mixings,
and hence the favored value of $BF$, shows very similar behavior
to the degenerate case (Figure$~$\ref{fig:Favored-BFs}, right plot).

In conclusion, the fully degenerate case with mixings is checked to
be allowed by the EW precision data. The favored values of the mixing
angles $\theta_{24}$ and $\theta_{34}$ tend to imply non-unity branching
fractions for the decays of the fourth generation quarks into the
third generation final states. The favored value of the $BF$ decreases
from about 90\% down to about 60\% with the fourth generation mass
increasing from 300 to 650$\,$GeV. Light Higgs scenarios with the
Higgs boson decaying to 4th generation neutrinos are allowed by the
EW precision data. Since mixing increases the $T$ parameter, for
heavy charged leptons low values of the mixing would be preferred.
However mixing can still play a role with behavior similar to the
fully-degenerate case.

\section{On the Fifth and Sixth Generations}

\begin{figure}[h]
\includegraphics[scale=0.4]{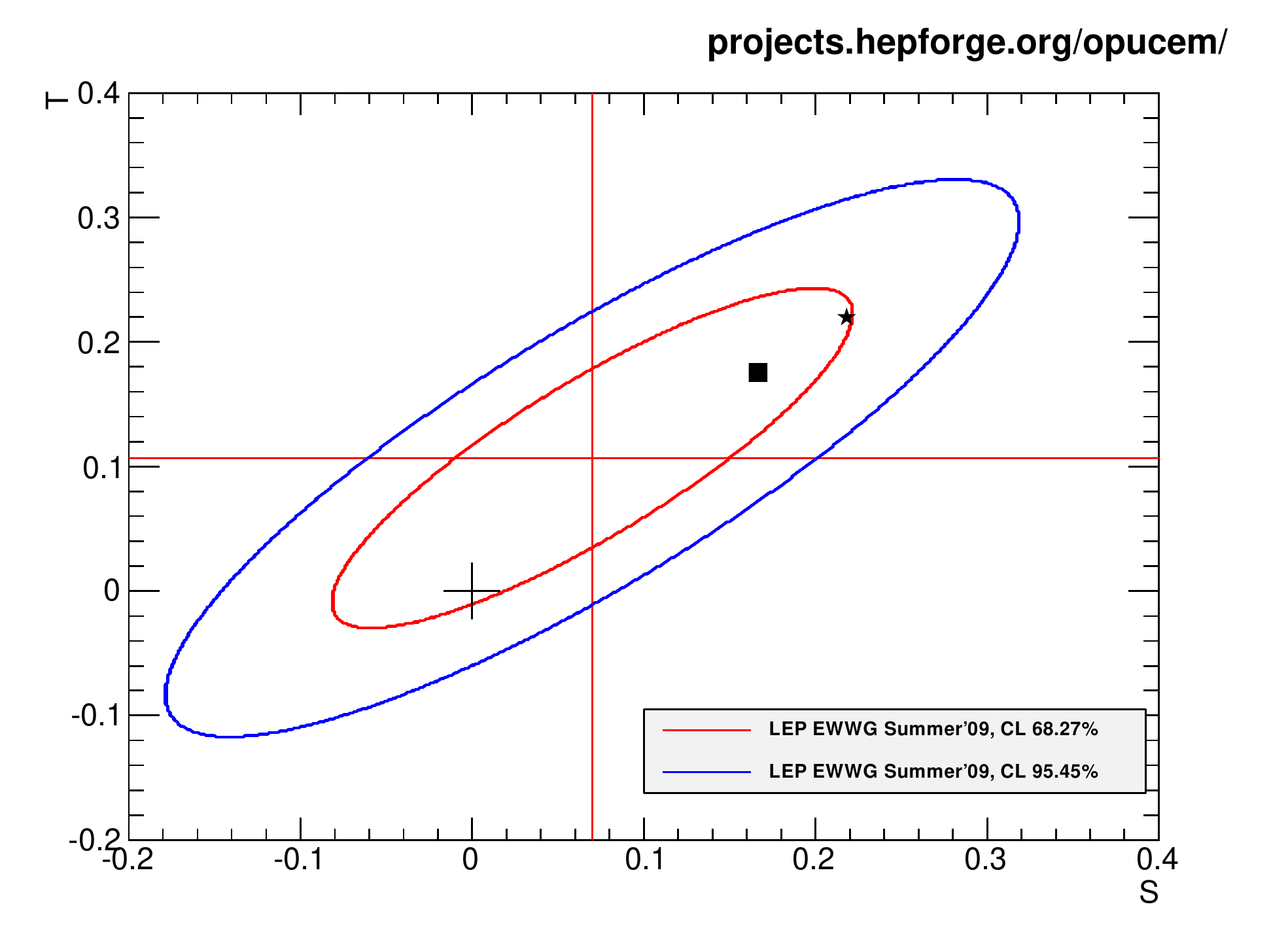}\includegraphics[scale=0.4]{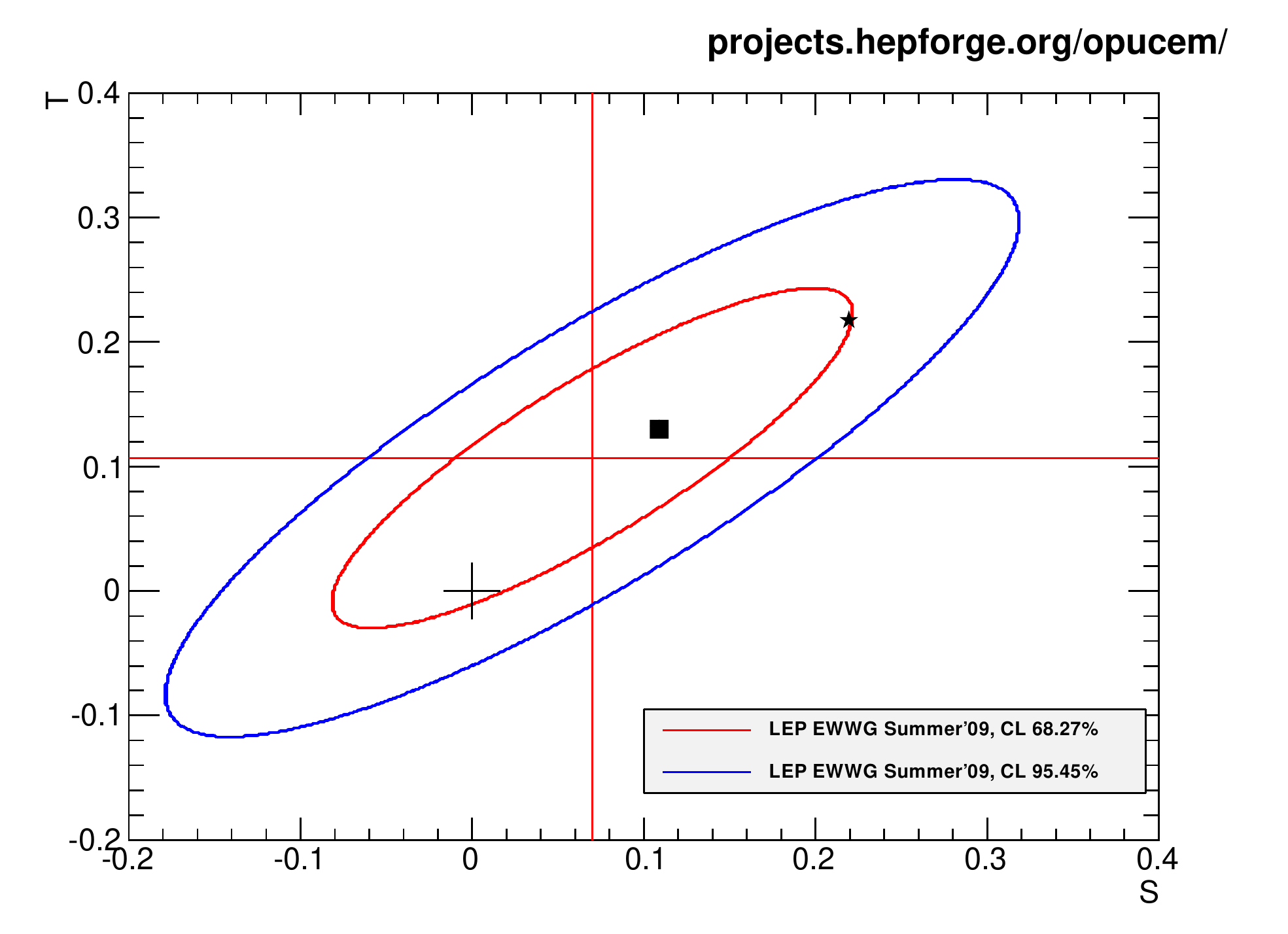}

\caption{SM3 and four SM5 benchmark points in $S-T$ plane with Dirac (left)
or Majorana (right) type neutrinos. The 1$\sigma$ and 2$\sigma$
error ellipses represent the 2009 results of the $U=0$ fit from the
LEP EWWG. The cross corresponds to the SM3 with $m_{H}=115$$\,$GeV.
In the left plot, the square(star) corresponds to the first(second)
SM5 point in Table$~$\ref{tab:Two-Dirac-type_SM5}. In the right
plot, the square(star) corresponds to the first(second) SM5 point
in Table \ref{tab:Two-Majorana-type_SM5}. \label{fig:SM3-and-two_SM5}}
\end{figure}

In this section, OPUCEM has been applied to the Standard Model with
5 and 6 generations (SM5 and SM6) in order to show that these are
not excluded by electroweak precision data. It is clear that the lack
of a Higgs signal from the LHC so far implies even stronger indirect
limits on a simple extension of the Standard Model with 5 or more
generations, and might have already excluded some parts of the available
parameter space. On the other hand, in certain region of the parameter
space, the addition of a 5th and 6th generation would reduce the effective
cross section in come channels even below the SM3 values \cite{higgs_sm5-6}.
Therefore, a study of these extra-generation cases is still interesting,
both as an input to beyond-SM model building with extra generations,
and as an exercise in understanding what the EW data might have indicated
had it been taken into account appropriately before. Keeping in mind
Autumn 2011 LHC results for Higgs boson searches, two values of Higgs
boson masses have been considered, namely $m_{H}=115$$\,$GeV and
$m_{H}=600$$\,$GeV. The masses of the extra fermions in SM5 with
Dirac neutrinos are presented in Table$~$\ref{tab:Two-Dirac-type_SM5}
for two example benchmark points. The corresponding points in the
$S-T$ plane are shown in Figure$~$\ref{fig:SM3-and-two_SM5} left
side together with the SM3 point. It can be seen that all three points
are inside the $1\sigma$ error ellipse.  For SM5 with Majorana type
neutrinos, two example benchmark points are listed in Table$~$\ref{tab:Two-Majorana-type_SM5},
with their locations on the $S$-$T$ plane shown on the right-hand
side of Figure$~$\ref{fig:SM3-and-two_SM5}.

\begin{table}[h]
\caption{Two SM5 benchmark points with Dirac type neutrinos. The first point
with $m_{H}=115$$\,$GeV and $|\sin\theta_{34}|=0.02$, leading to
$S=0.17$ and $T=0.17$ is on the left and the second point with $m_{H}=600$$\,$GeV
and $|\sin\theta_{34}|=0.02$, leading to $S=0.22$ and $T=0.22$
is on the right.\label{tab:Two-Dirac-type_SM5}}

\begin{tabular}{|c|c|c||c|c|}
\hline 
 & 4th family & 5th family & 4th family & 5th family\tabularnewline
\hline 
\hline 
$m_{U}$ (GeV) & $500$ & $550$ & $500$ & $550$\tabularnewline
\hline 
$m_{D}$ (GeV) & $500$ & $550$ & $500$ & $550$\tabularnewline
\hline 
$m_{\nu}$ (GeV) & $50$ & $50$ & $50$ & $50$\tabularnewline
\hline 
$m_{E}$ (GeV) & $120$ & $120$ & $150$ & $165$\tabularnewline
\hline 
\end{tabular}
\end{table}

\begin{table}[h]
\caption{Two SM5 benchmark points with Majorana type neutrinos. First point
with $m_{H}=115$$\,$GeV and $|\sin\theta_{34}|=0.07$, leading to
$S=0.11$ and $T=0.13$ is on the left and second point with $m_{H}=600$$\,$GeV
and $|\sin\theta_{34}|=0.07$, leading to $S=0.22$ and $T=0.22$
is on the right.\label{tab:Two-Majorana-type_SM5}}

\begin{tabular}{|c|c|c||c|c|}
\hline 
 & 4th family & 5th family & 4th family & 5th family\tabularnewline
\hline 
\hline 
$m_{U}$ (GeV) & $570$ & $580$ & $590$ & $590$\tabularnewline
\hline 
$m_{D}$ (GeV) & $500$ & $500$ & $500$ & $500$\tabularnewline
\hline 
$m_{\nu}$ (GeV) & $260$ & $45$ & $260$ & $45$\tabularnewline
\hline 
$m_{E}$ (GeV) & $590$ & $510$ & $590$ & $510$\tabularnewline
\hline 
$m_{N}$ (GeV) & $2550$ & $2900$ & $2550$ & $2900$\tabularnewline
\hline 
\end{tabular}
\end{table}

\begin{figure}[h]
\includegraphics[scale=0.4]{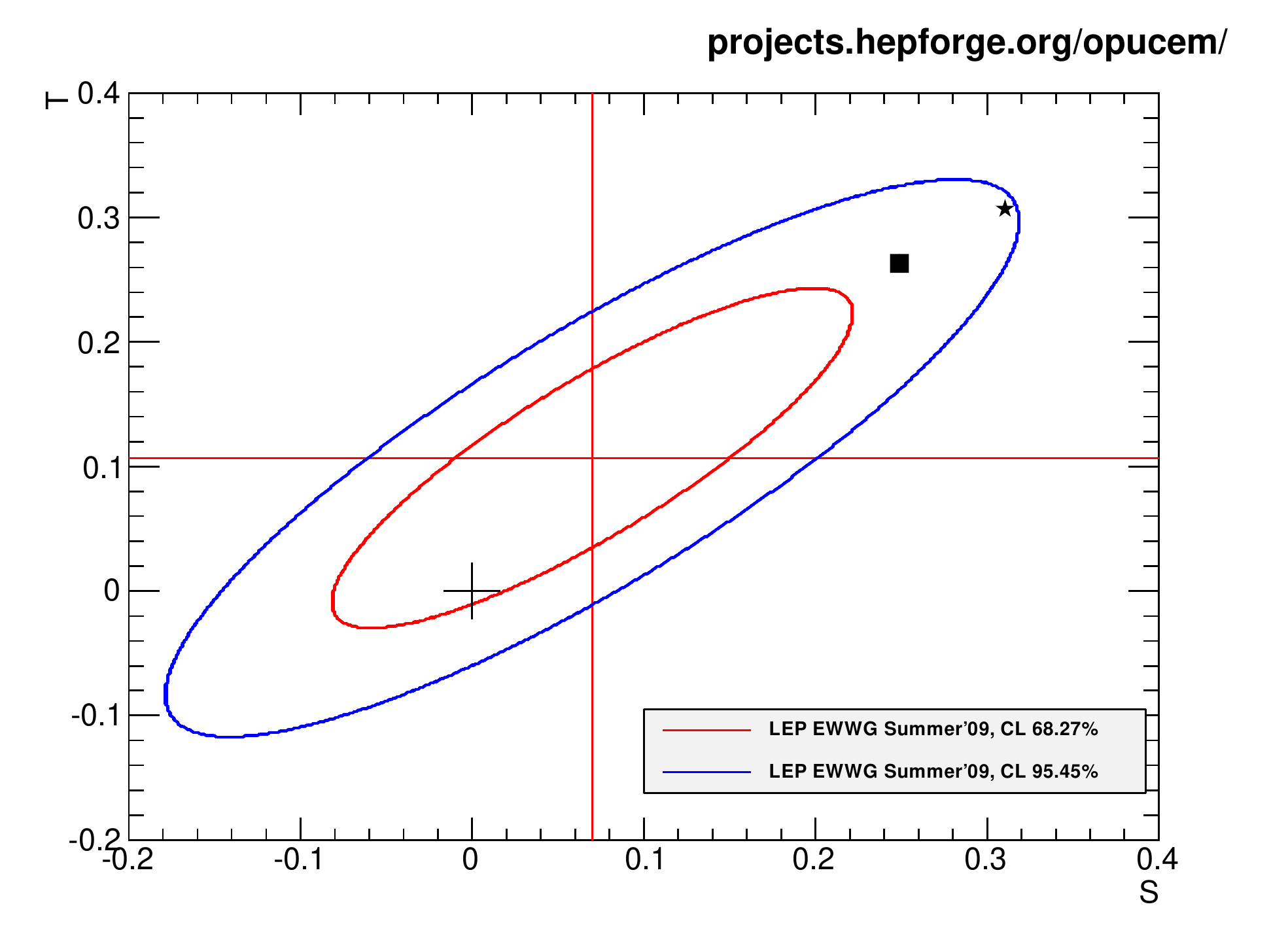}\includegraphics[scale=0.4]{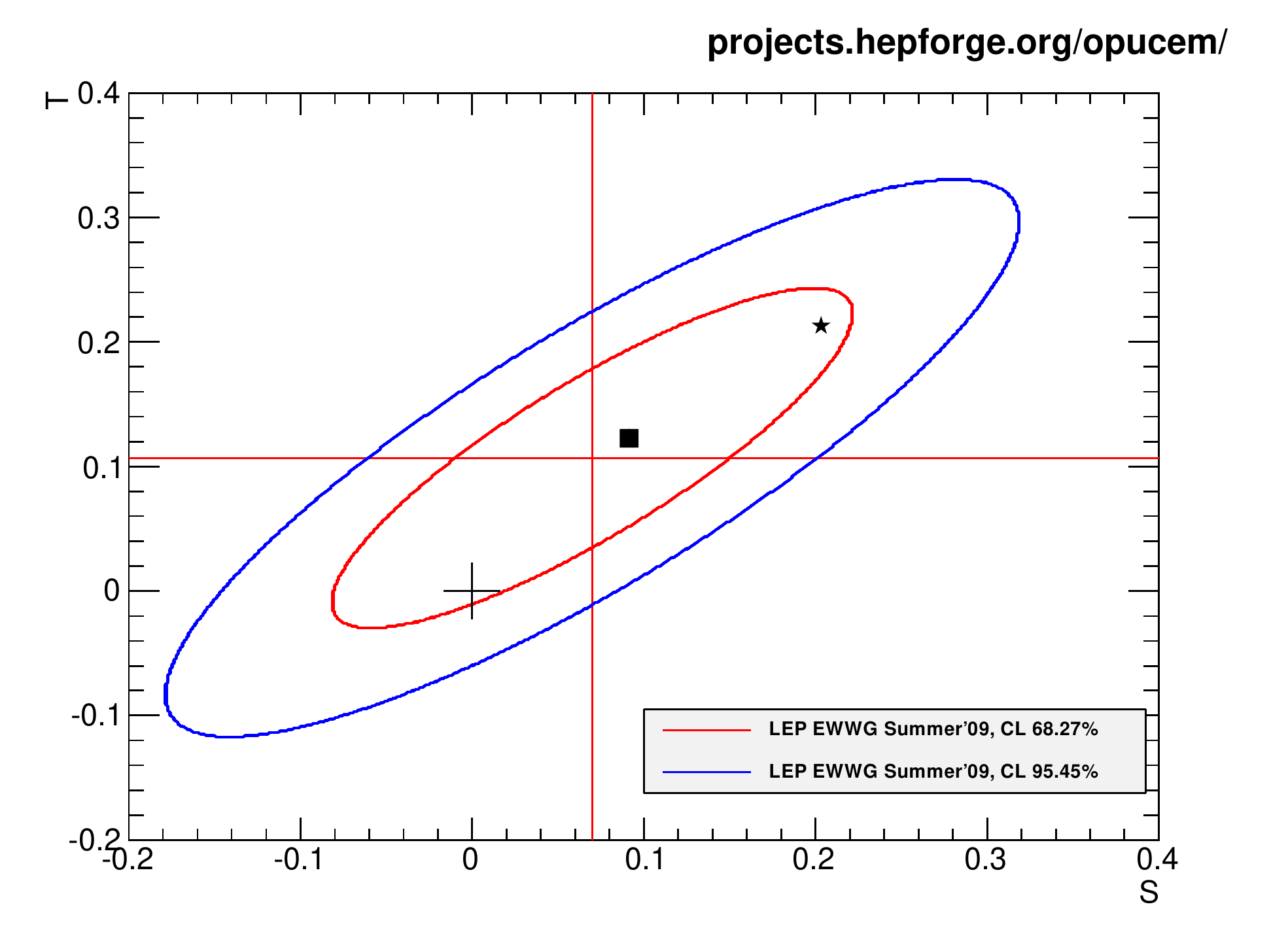}

\caption{SM3 and four SM6 benchmark points with Dirac (left) and Majorana(right)
type neutrinos. The 1$\sigma$ and 2$\sigma$ error ellipses represent
the 2009 results of the $U=0$ fit from the LEP EWWG. The cross corresponds
to the SM3 with $m_{h}=115$ GeV. In the left plot, the square(star)
corresponds to first(second) SM6 point in Table$~$\ref{tab:Two-Dirac-type_SM6}.
In the right plot, the square(star) corresponds to first(second) SM6
point in Table$~$\ref{tab:Two-Majorana-type_SM6}.\label{fig:SM3_and_SM6}}
\end{figure}

The SM6 with both Dirac and Majorana neutrinos has also been investigated.
Table$~$\ref{tab:Two-Dirac-type_SM6} contains two example sets of
mass values for the additional generations with Dirac type neutrinos.
Corresponding points in $S-T$ plane are shown in Figure$~$\ref{fig:SM3_and_SM6}
left side. It is seen that although the SM6 points lie further away
from the center of the ellipse than the SM3 point, they are still
within 2$\sigma$ error ellipse, thus not ruled out by the EW data.
Similarly, two example sets of mass values for SM6 with Majorana type
neutrinos are presented in Table$~$\ref{tab:Two-Majorana-type_SM6}
and the corresponding points in $S-T$ plane are shown in Figure$~$\ref{fig:SM3_and_SM6}
right side. One can see that these points are within 1$\sigma$ error
ellipse. It is worth noting that the SM6 points with $m_{H}=115$
GeV is significantly closer to central value than the SM3 point.

\begin{table}[h]
\caption{Two SM6 benchmark points with Dirac type neutrinos. The first point
with $m_{H}=115$$\,$GeV and $|\sin\theta_{34}|=0.00$, leading to
$S=0.25$ and $T=0.26$ is on the left and the second point with $m_{H}=600$$\,$GeV
and $|\sin\theta_{34}|=0.01$, leading to $S=0.31$ and $T=0.31$
is on the right. \label{tab:Two-Dirac-type_SM6}}

\begin{tabular}{|c|c|c|c||c|c|c|}
\hline 
 & 4th family & 5th family & 6th family & 4th family & 5th family & 6th family\tabularnewline
\hline 
\hline 
$m_{U}$ (GeV) & $530$ & $550$ & $580$ & $580$ & $580$ & $590$\tabularnewline
\hline 
$m_{D}$ (GeV) & $520$ & $550$ & $580$ & $580$ & $580$ & $590$\tabularnewline
\hline 
$m_{\nu}$ (GeV) & $50$ & $50$ & $50$ & $50$ & $50$ & $50$\tabularnewline
\hline 
$m_{E}$ (GeV) & $110$ & $130$ & $120$ & $110$ & $140$ & $180$\tabularnewline
\hline 
\end{tabular}
\end{table}

\begin{table}[h]
\caption{Two SM6 benchmark points with Majorana type neutrinos. The first point
with $m_{H}=115$$\,$GeV and $|\sin\theta_{34}|=0.01$, leading to
$S=0.09$ and $T=0.12$ is on the left and the second point with $m_{H}=600$$\,$GeV
and $|\sin\theta|=0.05$, leading to $S=0.20$ and $T=0.21$ is on
the right.\label{tab:Two-Majorana-type_SM6}}

\begin{tabular}{|c|c|c|c||c|c|c|}
\hline 
 & 4th family & 5th family & 6th family & 4th family & 5th family & 6th family\tabularnewline
\hline 
\hline 
$m_{U}$ (GeV) & $560$ & $570$ & $570$ & $580$ & $580$ & $580$\tabularnewline
\hline 
$m_{D}$ (GeV) & $500$ & $500$ & $500$ & $500$ & $500$ & $510$\tabularnewline
\hline 
$m_{\nu}$ (GeV) & $45$ & $45$ & $260$ & $45$ & $45$ & $260$\tabularnewline
\hline 
$m_{E}$ (GeV) & $410$ & $480$ & $550$ & $410$ & $480$ & $550$\tabularnewline
\hline 
$m_{N}$ (GeV) & $2400$ & $2600$ & $3500$ & $2200$ & $2500$ & $3500$\tabularnewline
\hline 
\end{tabular}
\end{table}

\section{Conclusions}

Using the enhancements to the OPUCEM library and associated tools,
it is shown that given the Autumn 2011 direct search results from
the LHC experiments, a sequential fourth family is still consistent
with the electroweak precision data. The statement is valid for both
cases with Dirac and Majorana neutrinos and also for both light and
heavy Higgs bosons. It should also be noted that the EW data still
prefers a light Higgs for both Dirac and Majorana type fourth generation.
Consideration of the mixing with light generation quarks enlarges
the available parameter space, making even the fully degenerate mass
scenario compatible with the EW precision data. Such a mixing also
reduces the current SM4 quark mass limits, as the favored \emph{BF}
to the third generation is found to be less than 100 percent. Finally,
analysis with OPUCEM shows that a fifth or even a sixth sequential
generation has been in the realm of the possible, with a $\chi^{2}$
lower than the three-generation SM in some example benchmark cases.
In conclusion, it is worth highlighting that the only way to discover
or completely rule out models with additional generations relies on
further data collection at the LHC and its analysis.
\begin{acknowledgments}
We would like to thank Saleh Sultansoy for many helpful conversations
and discussions. M. \c{S}ahin's work is supported by TUBITAK BIDEB-2218
grant. G. Unel's work is supported in part by U.S. Department of Energy
Grant DE FG0291ER40679. The collaboration leading to this paper has
partially been possible thanks to funds provided by the Bo\u{g}azi\c{c}i
University Foundation.\end{acknowledgments}

\end{document}